\documentclass[peerreview,a4paper]{IEEEtran}

\usepackage[utf8]{inputenc} 
\usepackage[T1]{fontenc}
\usepackage{url}
\usepackage{ifthen}
\usepackage{amssymb}
\usepackage{cite}
\usepackage{color}
\usepackage[cmex10]{amsmath} 

\usepackage{theorem}
\usepackage{bbm}

\newtheorem{thm}{\bf Theorem}
\newtheorem{cor}{Corollary}
\newtheorem{lem}{\bf Lemma}
\newtheorem{definition}{\bf Definition}
\newtheorem{rem}{\bf Remark}

\def\e{\mathbb{E}}
\def\p{\mathbb{P}}
\def\bx{\mathbf{x}}
\def\bX{\mathbf{X}}
\def\by{\mathbf{y}}
\def\bY{\mathbf{Y}}

\begin{document}
\allowdisplaybreaks
\title{Almost Exact Analysis of Soft Covering Lemma via Large Deviation} 


\author{%
  \IEEEauthorblockN{Mohammad H. Yassaee\\ School of Mathematics, Institute for Research in Fundamental Sciences (IPM)\\
                    Tehran, Iran\\
                    }
  }

\maketitle

\begin{abstract}
  This paper investigates the soft covering lemma under both the relative entropy and the total variation distance as the measures of deviation. The exact order of the expected deviation of the random i.i.d. code for the soft covering problem problem, is determined. The proof technique used in this paper  significantly differs from the previous techniques for deriving exact exponent of the soft covering lemma. The achievability of the exact order follows from applying the change of measure trick (which has been broadly used in the large deviation) to the known one-shot bounds in the literature. For the ensemble converse, some new inequalities of independent interest derived and then the change of measure trick is applied again. The exact order of the total variation distance is similar to the exact order of the error probability, thus it adds another duality between the channel coding and soft covering. Finally, The results of this paper are valid for any memoryless channels, not only  channels with finite alphabets.
  
  \end{abstract}


\section{Introduction}
There exist two  commonly used different (but dual) approaches for the investigating of the error exponent in the literature, which are closely related to the approaches used in the theory of large deviation \cite{LD,ld98};
\begin{enumerate}
\item The first one is based on the method of types. This approach is closely related to Sanov's approach\footnote{The general Sanov's approach is applicable to any random variable. Here by Sanov, we mean the approach used for discrete random variables which is also based on the method of types. }  in the large deviation (LD) theory. A comprehensive exposition of this approach for the basic problems can be found in \cite{csiszar:book}. 
\item Gallager's approach \cite{gallager:book}, which is similar in its spirit to the Cr\` amer approach in the LD.\end{enumerate}

While Sanov's method is more general than the Cramer's one, the latter has the advantage of being strong enough to find the exact order of the desired probability (such as the probability of deviating from the zero of the sum of independent r.v.'s). This observation was made by Bahadur and Rao, see \cite{LD}. In the same way, the exact order of random coding bound has been recently derived in \cite{yucel, scarlett2014mismatched}, using an approach related to Bahadur-Rao.

Soft covering lemma (also known as channel resolvability) \cite{han1993approximation,cs-cuff,wyner1975common} is another basic problem which has many applications such as secrecy problems, simulation  of channels, etc. Further, it is somehow the dual to the channel coding problem. Recently, the exact exponent of the soft covering lemma under different measures of closeness has been derived in \cite{parizi, YuTan,yagli}. Although the techniques used in these papers are different, all are based on the method of types and thus the results are limited to channels with finite alphabets. This motivates us to investigate other techniques similar to those used by Gallager and Cramer. Fortunately, such approach gives us  a new proof which is not only a different proof of the exact exponent that is valid for any channel with some regularity condition, but  leads to the exact order of the soft covering lemma.

The outline of the paper is as follows: after  stating the problem in section \ref{sec:notation}, we first present a one-shot upper bound for soft covering problem using  the total variation distance in Subsection \ref{subsec}, then we present our main results on the exact order of soft covering lemma using both the total variation distance and relative entropy in the rest of Section \ref{sec:ex-soft}. Section \ref{sec:analysis} is devoted to the ensemble converse proof of the results, while Section \ref{sec:achievability} is devoted to the achievability proof.

\section{Notations and Definitions}\label{sec:notation}
We follow closely the notation of Verd{\'u}'s book, \cite{IT}, with the exception of using Boldface letters to denote  vectors (e.g. $\bx=(x_1,\cdots,x_n)$). Throughout the the paper, the base of  the $\log$ and $\exp$ is $\mathrm{e}$. Also, we use the asymptotic notations such as $O(.),\Theta(.),\Omega(.)$ in the paper.

\begin{definition}{\bf Relative information}. \it  Given two measures $P$ and $Q$ on the same probability space such that $P \ll Q$, the relative information $\imath_{P||Q}$ is defined as
\[
\imath _ { P \| Q } ( x ) = \log \frac { \mathrm { d } P } { \mathrm { d } Q } ( x ).\]

\end{definition} 
\begin{definition}{\bf Information density.} \it Given a joint distribution $P_{XY}$, the information density  is defined as $\imath_{X;Y}(x,y)\triangleq \imath_{P_{XY}||P_X\times P_Y}(x,y).$ Throughout of the paper, we usually omit the subscript,  whenever it is clear from the context.
	
\end{definition} 
\begin{definition}\it
	For two distributions $P$ and $Q$ such that $P\ll Q$, the relative entropy and the total variation (TV) distance are defined as follows,
	\begin{align}
		D(P||Q)&:=\e[\imath_{P||Q}(X)]\\
		\|P-Q\|&:=\e[[\exp(\imath_{P||Q}(\overline{X}))-1]_+]=\frac{1}{2}\e[|\exp(\imath_{P||Q}(\overline{X}))-1|]
	\end{align}
	where $X\sim P$ , $\overline{X}\sim Q$ and $[x]_+\triangleq\max\{0,x\}$.
\end{definition}
\begin{definition}\it
	Given $P_{XY}$, the $\alpha$-mutual information  $I_{\alpha}(X;Y)$ \cite{IT} is defined by  $$I_{\alpha}(X;Y)=\frac{\alpha}{\alpha-1}\log\e[\e^{\frac{1}{\alpha}}[\exp(\alpha\imath_{X;Y}(X;\widetilde{Y}))|\widetilde{Y}]],$$ where   $(X,\widetilde{Y})\sim P_X P_Y$.
\end{definition}
\subsection{Problem Statement}  
Let $\mathcal{C}=\{X(k)\}_{k=1}^\mathsf{M}$ be a random codebook, in which its codewords are generated according to $P_X$. Given a channel $P_{Y|X}$, the output distribution $\mathsf{P}_Y$ induced by selecting uniformly an index $k$  from  $[1:\mathsf{M}]$ and then transmitting $X(k)$ through the channel, is
\[
\mathsf{P}_Y(.):=\frac{1}{\mathsf{M}}\sum_{k=1}^{\mathsf{M}}P_{Y|X}(.|X(k)).
\]
We are interested to evaluate the closeness of the induced distribution $\mathsf{P}_Y$ to $P_Y$, where $P_X\rightarrow P_{Y|X}\rightarrow P_Y$. We use the relative entropy and total variation distance to measure the closeness.  
\section{ Exact  Soft Covering Order}\label{sec:ex-soft}
\subsection{Gallager Type one-shot upper bound on TV-distance}\label{subsec}
We begin the investigation of soft covering problem by stating a Gallager type  upper bound on the TV-distance in the one-shot regime.
\begin{thm}\label{thm:TV-one-shot}\it
The TV-distance between the induced distribution $\mathsf{P}_Y$ and the desired distribution $P_Y$, is upper bounded by
	\begin{align}
	\e[\|\mathsf{P}_{Y}-P_Y\|]&\leq \dfrac{3}{2}\min_{0\le\rho\le\frac{1}{2}}~\mathsf{M}^{-\rho}.\exp\left(\rho I_{\frac{1}{1-\rho}}(X;Y)\right)
\label{eq:os-tv-e}
\end{align}
\end{thm}
\vskip 2mm
\begin{rem}\it
The one-shot bound \eqref{eq:os-tv-e}  readily implies that  the exact exponent of the soft covering lemma  for the i.i.d. codebook \cite[Theorem 1]{yagli} and for the constant-composition codebook \cite[Theorem 2]{yagli} is achievable, respectively. This follows by,
\begin{itemize}
	\item i.i.d. codebook. In this case, the codewords are drawn from $P_X^{\otimes n}$ and the channel $P_{Y^n|X^n}=\prod P_{Y|X}$ is memoryless. In this setting the assertion of \cite[Theorem 1]{yagli} follows from the identity  $I_{\frac{1}{1-\rho}}(\bX;\bY)=nI_{\frac{1}{1-\rho}}(X;Y)$.
	\item Constant composition codebook. In this case, the codewords are drawn uniformly from the set of all $\bx$ with the same type $P_X$, where $P_X$ is an $n$-type. In this setting, the assertion \cite[Theorem 2]{yagli} follows from the inequality  $I_{\frac{1}{1-\rho}}(\bX;\bY)\le nI^c_{\frac{1}{1-\rho}}(X;Y)$, where the Csiszar's $\alpha$-mutual information $I^c_{\alpha}(X;Y)$ is defined as
	\[
	I^c_{\alpha}(X;Y)=\inf_{Q_Y}\e[D_{\alpha}(P_{Y|X=X'}||Q_Y)]
	\]
	where $X'\sim P_X$.
\end{itemize}
Moreover, this exponent is achievable for any memoryless channel (not only the finite discrete memoryless one) with the assumption that the r.h.s. of \eqref{eq:os-tv-e} is finite. 
\end{rem}
\begin{rem}In a recent work \cite{MBAYF} , Mojahedian, et. al. consider wiretap channel $P_{YZ|X}$ and derive a lower bound bound on the exponent of the TV-distance between the joint distribution $P_{M\mathbf{Z}}$ of message $M$ and eavesdropper's observation $\mathbf{Z}$ and the product distribution $P_MP_{\mathbf{Z}}$ in term of Csiszar-$\alpha$-mutual information $I_{\alpha}(X;Z)$ (with the same exponent as \eqref{eq:os-tv-e} for the channel $P_{Z|X}$), using a completely different proof.\end{rem}
%
\begin{rem}\it Duality between Gallager's bound for channel coding and the exponent of the soft covering. The expression of 
	 \eqref{eq:os-tv-e} is the same as the  Gallager's one \cite[Eq. 78]{verdu2015alpha} for the channel coding with the exception that $\rho$ is replaced by $-\rho$.
\end{rem}
\vskip 2mm
\begin{IEEEproof}
The proof follows from the one-shot bound in \cite[Corollary VII.2]{cs-cuff} with a simple modification. The \cite[inequality (106)]{cs-cuff} asserts that 
\begin{align}
&	\e[\|\mathsf{P}_{Y}-P_Y\|]\leq \mathbb{P}[\mathcal{F}^c]\nonumber\\&~~~+\frac{1}{2}\e\left[\sqrt{\mathsf{M}^{-1}\e[\exp(\imath(X;Y))\mathbbm{1}\{(X,Y)\in\mathcal{F}\}|Y]}\right]\label{eq:os-tv}
\end{align} 
where $(X,Y)\sim P_{XY}$ and $\mathcal{F}$ is an arbitrary event\footnote{Cuff \cite{cs-cuff} only considered specific event $\mathcal{F}$ that gives a simple upper bound on the second term in \eqref{eq:os-tv}. However the  analysis is valid for any event $\mathcal{F}$.
 }. For any $0\leq \lambda\leq 1$, define 	
 \begin{align}
&	\mathcal{F}:=\Big\{(x,y):\exp(\imath(x;y))\nonumber\\
&~~~~~~~~~~~\leq \left(\mathsf{M}\e[\exp(\lambda\imath(X;Y))|Y=y]\right)^{\frac{1}{1+\lambda}}\triangleq \kappa_{\lambda}\Big \}
\end{align}
For a given $Y=y$, we have

\begin{align}
	\mathbb{P}[\mathcal{F}^c|Y=y]\leq \mathsf{M}^{-\frac{\lambda}{1+\lambda}}\e^{\frac{1}{1+\lambda}}[\exp(\lambda\imath(X;Y))|Y=y]\label{eq:tv-atyp}
\end{align}
where the inequality follows from the Markov inequality. Next, consider
\begin{align}
&\sqrt{\mathsf{M}^{-1}\e[\exp(\imath(X;Y))|Y=y]\mathbbm{1}\{(X,Y)\in\mathcal{F}\}}\nonumber\\
&\le\sqrt{\mathsf{M}^{-1}\kappa_\lambda^{1-\lambda}\e[\exp(\lambda\imath(X;Y))|Y=y]}\\
& = \mathsf{M}^{-\frac{\lambda}{1+\lambda}}\e^{\frac{1}{1+\lambda}}[\exp(\lambda\imath(X;Y))|Y=y]\label{eq:tv-sqrt}
\end{align}
where we used the definition of $\mathcal{F}$ and $\kappa_\lambda$.
  Observe that $$\e[\exp(\lambda\imath(X;Y))|Y=y]=\e[\exp((1+\lambda)\imath(X;y))]$$
by change of measure argument. Using this fact,  substituting  \eqref{eq:tv-atyp} and \eqref{eq:tv-sqrt} in \eqref{eq:os-tv} and setting $\rho=\frac{\lambda}{1+\lambda}$ imply \eqref{eq:os-tv-e}.
\end{IEEEproof}

\subsection{Exact order of Soft covering under relative entropy}
In the rest of the paper, we consider the soft covering problem for the memoryless channel $P_{Y|X}$ in the $n$-shot regime, with the codebook $\mathcal{C}$ consisting of $\mathsf{M}_n=\exp(nR)$ codewords such that the codewords are generated according to the i.i.d.\ distribution $P^{\otimes n}_X:=\prod_{k=1}^n P_X$. Here we denote the induced distribution with $\mathsf{P}_{Y^n}$.
\begin{thm}\it\label{thm:KL}
Suppose that $R>I(X;Y)>0$. Further, assume that the moment generating function of $\imath(X;Y)$ is finite in the neighborhood of origin, that is $\e[\exp(\tau\imath(X;Y)]<\infty$ in the  neighborhood of origin.  Let
	\begin{equation}
	\tau^*=\arg\max_{0\leq\tau\leq 1} \tau R-\log\e[\exp(\tau\imath_{X;Y}(X;Y))].\label{eq:tau-defn}
\end{equation}
Then
\begin{align}
&\e\left[D(\mathsf{P}_{Y^n}||P^{\otimes n}_Y)\right]\nonumber\\&=\left\{\begin{array}{ll}
\Theta\left(\frac{\exp(-n\tau^*R)}{\sqrt{n}}\e^n[\exp(\tau^*\imath(X;Y))]\right)&\tau^*<1
\\
\Theta\left({\exp(-nR)}\e^n[\exp(\imath(X;Y))]\right)& \tau^*=1\end{array}\right. \label{eq:KL-Exact}
\end{align}
\end{thm}
\subsection{Exact order of Soft covering under TV distance}
\begin{definition}\it
	A channel $P_{Y|X}$ is said singular, if $\mathrm{Var}[\imath(X;Y)|Y]=0$, almost surly w.r.t. $P_Y$.\footnote{For the discrete channel, this definition is equivalent to the definition of singular channel in \cite[Definition 1]{yucel}.} Otherwise, the channel is non-singular.
\end{definition}
\begin{thm}\it\label{thm:exact-tv}
	Suppose that $R>I(X;Y)>0$ and $\e[\exp(\tau\imath(X;Y)]<\infty$ in the  neighborhood of origin. Let 
	\begin{align}
		\rho^*=\arg&\max_{0\leq\rho\leq \frac{1}{2}} \rho (R-I_{\frac{1}{1-\rho}}(X;Y))
		\label{eq:rho-defn}
	\end{align}
	 To state the exact order, we should distinguish between singular and non-singular channels. We have
	\begin{align}
&\e\left[\|\mathsf{P}_{Y^n}-P^{\otimes n}_Y\|\right]\nonumber\\&=\left\{\begin{array}{ll}
\Theta\Big({{n^{-\frac{\beta^*}{2}}}}\exp(-{n\rho^*}(R-I_{\frac{1}{1-\rho^*}}(X;Y))\Big)&\rho^*<\frac{1}{2}
\\
\Theta\left(\exp(-\frac{n}{2}(R-I_{2}(X;Y))\right)& \rho^*=\frac{1}{2}
\end{array}\right.
\label{eq:TV-Exact}
\end{align}
where $\beta^*=1-\rho^*$, for the  non-singular channels and $\beta^*=1$ for the singular channels.
\begin{rem}{\it Duality.} \it Again, the expression of the \eqref{eq:TV-Exact} is similar to the expression of the exact order of random coding bound for the channel coding \cite{yucel}, except that $\rho$ is replaced by $-\rho$. While the expressions are similar, the proofs are quite different. 
	\end{rem}
\end{thm}
\section{Exact analysis for Ensemble Converse}\label{sec:analysis}
In this section, we present the ensemble converse proof of the Theorem \ref{thm:KL} and Theorem \ref{thm:exact-tv}. The proof is divided to four main steps. To make the analysis concise, we utilize  the idea of Poissonizating the problem, which have been used in \cite{yagli} to eliminate the correlation between weakly dependent r.v's. Using the concentration of the Poisson r.v. around its mean, we show that the exact order of the relative entropy and TV-distance after Poissonization is the same the main problems for the fixed rate. So it suffices to find the exact oreder of the Poissonizated problem. To evaluate the exact order of Poissonizated problem, we use the thinning property of certain Poisson random sum to find lower bounds on the desired parameters. Next, we further lower bounded  bounds in terms of moments of certain r.v. We present these steps in parallel for both the relative entropy and TV-distance. Then, we continue the analysis separately for these two cases, although the main trick is the same.  We use the change of measure trick in the same way as the one used in the converse proof of the Bahadur-Rao \cite{LD} theorem (i.e. the exact order of the probability of the deviating of a sum from the mean) to find the exact order.    
\subsection{Poissonization}

By Poissonization, we assume that the number of codewords is not fixed, but is a Poisson random variable, with the mean close to the size of the codebook. More precisely, we assume that the Poisson-codebook is $\{\bX(k)\}_{k\in\mathbb{N}}$, where the codewords are generated according to $P^{\otimes n}_X$. Further, we assume that \underline{$M$ is a Poisson r.v.} with {\it mean} $\mu_n=2\exp(nR)$. 

Let $\mathsf{L}_m$ and $\mathsf{V}_m$ be the average of the relative entropy and the TV distance, respectively, when the number of codewords is $m$, that is 
\begin{IEEEeqnarray}{rcl}
\mathsf{L}_{m}=\e\left[ D\left( \mathsf{P}_{\mathbf{Y}}^{(m)}||P^{\otimes n}_{Y}\right) \right] \label{eq:LM-dfn}\\
\mathsf{V}_{m}=\e\left[ \left\| \mathsf{P}_{\mathbf{Y}}^{(m)}-P^{\otimes n}_{Y}\right\| \right]\label{eq:VM-dfn}
\end{IEEEeqnarray}
where
 \(
\mathsf{P}_{\mathbf{Y}}^{(m)}(.):=\frac {1}{m}\sum ^{m}_{k=1}P_{\mathbf{Y}|\mathbf{X}}\left( .|\mathbf{X}\left( k\right) \right) 
\). 
 We will show that $\e[\mathsf{L}_M]$ is a good approximation for $\mathsf{L}_{\exp(nR)}$. Also $\e[\mathsf{V}_M]$ is a good approximation for $\mathsf{V}_{\exp(nR)}$. More precisely, we have,
\begin{lem}\label{le:1}
\begin{align}
	\mathsf{L}_{\exp(nR) }&\ge 
\mathbb{E}\left[ \mathsf{L}_{M}\right] - nI\left( X;Y\right) \varepsilon_{\frac {1}{2}}^{\mu _{n}}\\
\mathsf{V}_{\exp(nR) }&\ge 
\mathbb{E}\left[ \mathsf{V}_{M}\right] -  \varepsilon_{\frac {1}{2}}^{\mu _{n}}
\end{align}
where $\varepsilon_{\frac{1}{2}}=\sqrt{2}{\mathrm{e}^{-\frac{1}{2}}}<1$.
\end{lem}
\begin{IEEEproof}
By Lemma \ref{le:mono-f-divergence} in the Appendix \ref{apx:mono-f-divergence} , the sequence $\{\mathsf{L}_m\}$ is a decreasing sequence of $m$.
 Further $\mathsf{L}_1=\e[D(P_{\bY|\bX=\bX_1}||P_\bY)]=nI(X;Y)$, because $\bX_1\sim P_X^{\otimes n}$. Thus,
 \begin{align}
\e\left[ \mathsf{L}_{M}\right] &\leq \mathsf{L}_1\p[M< \exp(nR)]+\mathsf{L}_{\exp(nR)}\p[M\ge \exp(nR)]\\
&\le nI\left( X;Y\right) \varepsilon_{\frac {1}{2}}^{\mu _{n}}+\mathsf{L}_{\exp \left( nR\right) }.
\end{align}	
where the last inequality follows from \cite[Theorem 5.4]{mitz}.

Similarly, the sequence $\{\mathsf{V}_m\}$ is decreasing by Lemma \ref{le:mono-f-divergence}, thus
 \begin{align}
\e\left[ \mathsf{V}_{M}\right] &\leq \mathsf{V}_1\p[M< \exp(nR)]+\mathsf{V}_{\exp(nR)}\p[M\ge \exp(nR)]\\
&\le  \varepsilon_{\frac {1}{2}}^{\mu _{n}}+\mathsf{V}_{\exp \left( nR\right) }.
\end{align}	
where we have used $\mathsf{V}_1\le 1$.
\end{IEEEproof}

Let $T$ be a random variable  defined by\footnote{It is worthy to note that the randomness in $T$ comes from the randomness of the codebook, poisson r.v. $M$ and the r.v. $\bY\sim P_Y^{\otimes n}$.}

\begin{equation}
	T=\frac {1}{\mu _{n}}\sum ^{M}_{k=1}\exp \left( \imath\left( \mathbf{X}(k);\mathbf{Y}\right) \right) . 
\end{equation}

\begin{lem}\label{le:2}
\begin{align}
\e\left[ \mathsf{L}_{M }\right] &\geq \frac {1}{2}\e\left[T\log  {T}\right] -\frac {1}{2\mu _{n}}\left( 1+\mathsf{L}_{2\mu _{n}}
\mu _{n}\varepsilon _{\frac{3}{2}}^{\mu _{n}}\right)\\
&= \frac {1}{2}\e\left[T\log  {T}\right] -O\left(\exp(-nR)\right)	\end{align}
	where $\varepsilon_{\frac{3}{2}}=\frac{\mathrm{e}^{.5}}{1.5^{1.5}}<1$. 	\end{lem}

\begin{lem}\label{le:TV2}
\begin{align}
\e\left[ \mathsf{V}_{M }\right] &\geq \frac {1}{4}\e\left[\Big|T-1\Big|\right] -\frac {1}{4\sqrt{\mu _{n}}}-\frac{1}{2}
\varepsilon _{\frac{3}{2}}^{\mu _{n}}\\
&= \frac {1}{4}\e\left[\Big|T-1\Big|\right] -O\left(\exp(-n\frac{R}{2})\right).	\end{align}
	
\end{lem}
The proofs are Lemma \ref{le:2} and Lemma \ref{le:TV2} are provided in the Appendix \ref{apx:relative-entropy} and Appendix \ref{apx:TV}, respectively.
Comparing Lemma \ref{le:1} with Lemmas \ref{le:2} and \ref{le:TV2}, we get
\begin{cor}\label{cor:both}
\begin{align}
		\mathsf{L}_{\exp(nR)}&\geq \frac {1}{2}\e\left[T\log  {T}\right] -O\left(\exp(-nR)\right),\label{eq:190}\\
		\mathsf{V}_{\exp(nR)}&\geq \frac {1}{4}\e\left[\Big|T-1\Big|\right] -O\left(\exp(-n\frac{R}{2})\right).\label{eq:TV-190}
\end{align}	
\end{cor}
\subsection{Negligibility of the O-terms in the Corollary \ref{cor:both}} 
We show that the O-terms in \eqref{eq:190} and \eqref{eq:TV-190} are negligible with respect to the exact expressions \eqref{eq:KL-Exact} and \eqref{eq:TV-Exact}, respectively. Thus, it is only required to prove the exact expressions are lower bounds for the expectation terms in the Corollary \ref{cor:both}.

Observe that the exact exponent in the exact order \eqref{eq:KL-Exact} is
\begin{align}
	\max_{0\le\tau\leq 1}\tau R-\log\e[\exp(\tau\imath_{X;Y}(X;Y))]&\leq \max_{0\le\tau\leq 1} \tau (R-I(X;Y))\\
	&<R 
\end{align}
where we used the Jensen inequality for the concave function $\log x$ and the assumption $I(X;Y)>0$. Thus the O-term $O\left(\exp(-nR)\right)$ is negligible w.r.t. the exact order \eqref{eq:KL-Exact}.

Similarly, the exact exponent in the exact order \eqref{eq:TV-Exact} is
\begin{align}
	\max_{0\leq\rho\leq \frac{1}{2}} \rho (R-I_{\frac{1}{1-\rho}}(X;Y))\leq\max_{0\leq\rho\leq \frac{1}{2}} \rho (R-I(X;Y))<\frac{R}{2},
\end{align}
where we used the fact that $I_s(X;Y)$ is an increasing function of $s$.

\subsection{Lower bounding using Tinning property of Poisson random sum}
Let $\mathcal{F}$ be an arbitrary event. To obtain a lower bound on $\e[T\log T]$ (for the TV- case, $\e[|T-1|]$), we split $T$ to two parts $T_1$ and $T_2$ defined below,
\begin{align}
 {T}_1&=\frac {1}{\mu _{n}}\sum ^{M}_{k=1}\exp \left( \imath\left( \mathbf{X}(k);\mathbf{Y}\right) \right) \mathbbm{1}\left\{ \left( \mathbf{X}\left( k\right) ,\mathbf{Y}\right) \in \mathcal{F}\right\} \\
 {T}_2&=\frac {1}{\mu _{n}}\sum ^{M}_{k=1}\exp \left( \imath\left( \mathbf{X}(k);\mathbf{Y}\right) \right) 1{\left\{ \left( \mathbf{X}\left( k\right) ,\mathbf{Y}\right) \notin \mathcal{F}\right\} }
\end{align}

It is clear that $T=T_1+T_2$. Further conditioned on any instance $\bY=\by$, the random variables $U_k:=\exp \left( \imath\left( \mathbf{X}(k);\mathbf{Y}\right) \right)$ are i.i.d. So  the tinning property of the Poisson random sum $\sum_{k=1}^M U_k$, (which is proved in the Appendix \ref{apx:Poisson-tinning}) shows that  $T_1$ and $T_2$ are independent given $\bY=\by$.

 Moreover, $$\e[T|\bY=\by]=\frac{1}{\mu_n}\e\left[\e\left[\sum_{k=1}^M\exp(\imath(\bX_k;\by))\right]\Big|M\right]=\frac{1}{\mu_n}\e\left[M\e\left[\exp(\imath(\bX_1;\by))\right]\right]=\frac{\e[M]}{\mu_n}=1.$$ Thus using the Jensen inequality for the convex function $f(x)=x\log x$, we have 
\begin{align}
\e[T\log T|\bY]&=\e[(T_1+T_2)\log (T_1+T_2)|\bY]\\
&\ge \e\Big[(T_1+\e[T_2|\mathbf{Y}])\log({T}_1+\e[{T}_2|\mathbf{Y}])|\bY\Big]\\
&= \e\Big[(T_1+1-\e[T_1|\mathbf{Y}])\log({T}_1+1-\e[{T}_1|\mathbf{Y}])|\bY\Big].\label{eq:22}	
\end{align}
Similarly the Jensen inequality for the convex function $f(x)=|x-1|$ implies
\begin{equation}
	\e\left[|T-1|\Big|\bY\right]\ge \e\Big[|T_1-\e[T_1|\mathbf{Y}]|\Big|\bY\Big].\label{eq:V22}
\end{equation}
\subsection{Useful bounds on $\e[U\log U]$ and $\e[|U-\e[U]|]$ in terms of the moments and their consequences}
The following lemma, which is of independent interest, plays the key role in  proving the converse for the relative entropy. Its proof is given in the Appendix \ref{apx:log}.

\begin{lem}\label{le:conv}
For  a positive random variable $U$   with $\e[U]=1$, we have 
\begin{equation}
	\e\left[ U\log U\right]\geq \dfrac {\e\left[ \left( U-1\right) ^{2}\right] ^{2}}{{2}\e\left[ \left( U-1\right) ^{2}\right] +\dfrac {2}{3}\e\left[ \left( U-1\right) ^{3}\right] }.
\end{equation}
\end{lem}

Also, the following lemma is the TV-counterpart of the previous lemma.
\begin{lem}\label{le:TV-conv}
For  any positive random variable $U$, we have 
\begin{equation}
	\e\left[ \big|U-\e[U]\big|\right]\geq \sqrt{\dfrac {\e\left[ \left( U-\e[U]\right) ^{2}\right] ^{3}}{\e\left[ \left( U-\e[U]\right) ^{4}\right]  }}.\label{eqn:TV-conv}
\end{equation}
\end{lem}
\begin{IEEEproof}For any r.v. $V$, we have
	\begin{align}
		\e[|V|]^{\frac{2}{3}}\e[V^4]^{\frac{1}{3}}\geq \e[V^2]\label{eq:Holder-00}
	\end{align}
	where we have used the Holder inequality. Setting $V\leftarrow U-\e[U]$ and rearranging \eqref{eq:Holder-00} yield \eqref{eqn:TV-conv}.
\end{IEEEproof}

Using Lemma \ref{le:conv}, we prove the following lemma in the Appendix \ref{apx:sub-kl}.
\begin{lem}\label{le:4} For any event $\mathcal{F}$,
\begin{align}
\e[T\log T]\ge&\frac{1}{4}\min\left\{ \frac{1}{\mu_n}\e\left[\exp(\imath(\mathbf{X};\mathbf{Y})\mathbbm{1}\{\mathbf{X},\mathbf{Y})\in\mathcal{F}\}\right]\right.\nonumber\\
&,\left. 3\dfrac{\e\left[\exp(\imath(\mathbf{X};\mathbf{Y})\mathbbm{1}\{\mathbf{X},\mathbf{Y})\in\mathcal{F}\}\right]^2}{\e\left[\exp(2\imath(\mathbf{X};\mathbf{Y})\mathbbm{1}\{\mathbf{X},\mathbf{Y})\in\mathcal{F}\}\right]} \right\}
\end{align}
where $(\bX,\bY)\sim P_{XY}^{\otimes n}$.	
\end{lem}

Also, using Lemma \ref{le:TV-conv}, we prove the following lemma in the Appendix \ref{apx:sub-TV}.

\begin{lem}\label{le:TV-4} For any event $\mathcal{F}$,
\begin{align}
\e\left[\big|T-1\big|\right]
&\ge\e\sqrt{\dfrac {1}{\dfrac{\e\left[\exp(3\imath(\mathbf{X};\mathbf{Y})\mathbbm{1}\{\mathbf{X},\mathbf{Y})\in\mathcal{F}\}|\bY\right]}{\e\left[\exp(\imath(\mathbf{X};\mathbf{Y})\mathbbm{1}\{\mathbf{X},\mathbf{Y})\in\mathcal{F}\}|\bY\right] ^{3}}+{3}\mu_n\e\left[\exp(\imath(\mathbf{X};\mathbf{Y})\mathbbm{1}\{\mathbf{X},\mathbf{Y})\in\mathcal{F}\}|\bY\right]^{-1} }}
\end{align}
where $(\bX,\bY)\sim P_{XY}^{\otimes n}$.	
\end{lem}

\subsection{Large deviation type analysis for Relative entropy}
To evaluate the bound in Lemma \ref{le:4}, we use the change of measure trick in the same spirit as the one used in the large deviation for proving Cramer theorem and its extension by Bahadur-Rao, see \cite{LD}. 

Define the tilted distribution $P_{X_*Y_*}$ via the following Radon-Nikodym derivative,
\begin{equation}
\dfrac{\mathrm{d}P_{X_*Y_*}}{\mathrm{d}P_{XY}}(x,y):=\dfrac{\exp(\tau^*\imath(x;y))}{\e[\exp(\tau^*\imath(X;Y))]} \triangleq \dfrac{\exp(\tau^*\imath(x;y))}{S}\label{eqn:tilted}
\end{equation}
where $\tau^*$ was defined in \eqref{eq:tau-defn}. We consider the cases $\tau^*<1$ and $\tau^*=1$, separately.

{\bf Case I:    $\tau^*<1$}. Differentiating the function inside \eqref{eq:tau-defn}  and equalling it to zero gives, 
\begin{align}
	R&=\dfrac{\e[\imath_{X;Y}(X;Y)\exp(\tau^*\imath_{X;Y}(X;Y))]}{\e[\exp(\tau^*\imath_{X;Y}(X;Y))]}=\e[\imath_{X;Y}(X_*;Y_*)]
\end{align}
Now set,
\begin{align}
&\mathcal{F}:=\{(\mathbf{x},\by):n\e[\imath_{X;Y}(X_*;Y_*)]\le\imath_{\mathbf{X};\mathbf{Y}}(\mathbf{x};\by)\nonumber\\
&\qquad\qquad\qquad\qquad\quad\qquad\le n\e[\imath_{X;Y}(X_*;Y_*)]+A\}
\end{align}
Here we choose the positive constant $A$  large enough such that $\p[(\bX_*,\bY_*)\in\mathcal{F}]\geq \frac{C}{\sqrt{n}}$ for some positive constant $C$, where $(\bX_*,\bY_*)\sim P_{X_*Y_*}^{\otimes n}$. The existence of such $A$ is guaranteed by the application of Berry-Essen CLT to the r.v. $\imath_{\bX;\bY}(\bX_*;\bY_*)=\sum_{i=1}^n \imath_{X;Y}(X_{*,i};Y_{*,i})$.
Then for $\tau^*<1$ , we have,
\begin{align}
&\e\left[\exp(\imath(\mathbf{X};\mathbf{Y}))\mathbbm{1}\{\mathbf{X},\mathbf{Y})\in\mathcal{F}\}\right]\nonumber\\
&~= S^n\e\left[\exp((1-\tau^*)\imath(\mathbf{X}_*;\mathbf{Y}_*))\mathbbm{1}\{\mathbf{X}_*,\mathbf{Y}_*)\in\mathcal{F}\}\right]\label{eq:ch35}\\
&~\geq S^n\exp(n(1-\tau^*)\e[\imath_{X;Y}(X_*;Y_*)])\nonumber\\
&\qquad\qquad\qquad\qquad\qquad\qquad\qquad\mathbb{P}\left[(\mathbf{X}_*,\mathbf{Y}_*)\in\mathcal{F}\right]\label{eq:ch36}\\
&~= S^n\exp(n(1-\tau^*)R)\mathbb{P}\left[(\mathbf{X}_*,\mathbf{Y}_*)\in\mathcal{F}\right]\label{eq:ch37}
\end{align}
where \eqref{eq:ch35} follows by change of measure using the definition of $P_{X_*Y_*}$ and \eqref{eq:ch36} follows from the definition of the event $\mathcal{F}$.

Similarly we have,
\begin{align}
&\e\left[\exp(2\imath(\mathbf{X};\mathbf{Y}))\mathbbm{1}\{\mathbf{X},\mathbf{Y})\in\mathcal{F}\}\right]\nonumber\\
&~= S^n\e\left[\exp((2-\tau^*)\imath(\mathbf{X}_*;\mathbf{Y}_*))\mathbbm{1}\{\mathbf{X}_*,\mathbf{Y}_*)\in\mathcal{F}\}\right]\\
&~\leq S^n\exp((2-\tau^*)(nR+A))\mathbb{P}\left[(\mathbf{X}_*,\mathbf{Y}_*)\in\mathcal{F}\right]\label{eq:ch39}
\end{align}
Substituting \eqref{eq:ch37} and \eqref{eq:ch39} Lemma \eqref{le:4}, implies that for some $C_1>0$, 
\begin{align}
	\e[T\log T]&\geq C_1 S^n\exp(-n\tau^*R)\p[(\bX_*,\bY_*)\in\mathcal{F}]\\
	&=\Omega\left(\frac{\exp(-n\tau^*R)}{\sqrt{n}}\e^n[\exp(\tau^*\imath(X;Y))]\right)\label{eq:410}
\end{align} 
Putting this in \eqref{eq:190} concludes the converse proof of Theorem \ref{thm:KL}.

{\bf Case II:    $\tau^*=1$}. Lemma \ref{le:apx-kl} in the Appendix \ref{apx:optimum} implies 
\begin{align}
	R&\ge\dfrac{\e[\imath_{X;Y}(X;Y)\exp(\imath_{X;Y}(X;Y))]}{\e[\exp(\imath_{X;Y}(X;Y))]}=\e[\imath_{X;Y}(X_*;Y_*)]\label{eqn:tilt-c}
\end{align}
where $P_{X_*Y_*}$ is defined by \eqref{eqn:tilted} with $\tau^*=1$. 
Now set,
\begin{align}
&\mathcal{F}:=\{(\mathbf{x},\by):\imath_{\mathbf{X};\mathbf{Y}}(\mathbf{x};\by)
\le n\e[\imath_{X;Y}(X_*;Y_*)]\}
\end{align}
We have,
\begin{align}
\e\left[\exp(\imath(\mathbf{X};\mathbf{Y}))\mathbbm{1}\{(\mathbf{X},\mathbf{Y})\in\mathcal{F}\}\right]
&= S^n\e\left[\mathbbm{1}\{\mathbf{X}_*,\mathbf{Y}_*)\in\mathcal{F}\}\right]\label{eq:ch-1-35}\\
& =S^n\p\left[(\mathbf{X}_*,\mathbf{Y}_*)\in\mathcal{F}\right]\label{eq:ch-1-37}\end{align}
where \eqref{eq:ch-1-35} follows by change of measure using the definition of $P_{X_*Y_*}$.

Similarly we have,
\begin{align}
&\e\left[\exp(2\imath(\mathbf{X};\mathbf{Y}))\mathbbm{1}\{\mathbf{X},\mathbf{Y})\in\mathcal{F}\}\right]\nonumber\\
&~= S^n\e\left[\exp(\imath(\mathbf{X}_*;\mathbf{Y}_*))\mathbbm{1}\{\mathbf{X}_*,\mathbf{Y}_*)\in\mathcal{F}\}\right]\\
&~\leq S^n\exp(n\e[\imath_{X;Y}(X_*;Y_*)])\mathbb{P}\left[(\mathbf{X}_*,\mathbf{Y}_*)\in\mathcal{F}\right]\label{eq:ch-1-39}
\end{align}
Substituting \eqref{eq:ch-1-37} and \eqref{eq:ch-1-39} in Lemma \eqref{le:4}, yields
\begin{align}
	\e[T\log T]&\ge\frac{1}{4}\min\left\{ \frac{1}{\mu_n}\e\left[\exp(\imath(\mathbf{X};\mathbf{Y})\mathbbm{1}\{\mathbf{X},\mathbf{Y})\in\mathcal{F}\}\right],3\dfrac{\e\left[\exp(\imath(\mathbf{X};\mathbf{Y})\mathbbm{1}\{\mathbf{X},\mathbf{Y})\in\mathcal{F}\}\right]^2}{\e\left[\exp(2\imath(\mathbf{X};\mathbf{Y})\mathbbm{1}\{\mathbf{X},\mathbf{Y})\in\mathcal{F}\}\right]} \right\}\\
	&\ge\frac{\mathbb{P}\left[(\mathbf{X}_*,\mathbf{Y}_*)\in\mathcal{F}\right]}{4}S^n\min\left\{ \frac{1}{2}\exp(-nR),3\exp(-n\e[\imath_{X;Y}(X_*;Y_*)]) \right\}\\
	&\ge\left(\frac{1}{16}+O\left(\frac{1}{\sqrt{n}}\right)\right)S^n \exp(-nR)
	\end{align} 
	where the last inequality follows from \eqref{eqn:tilt-c} and the Berry-Esseen approximation $\mathbb{P}[(\mathbf{X}_*,\mathbf{Y}_*)\in\mathcal{F}]=\frac{1}{2}+O(n^{-\frac{1}{2}})$.
Putting this in \eqref{eq:190} concludes the converse proof of Theorem \ref{thm:KL}.

\subsection{Large deviation type analysis for TV-distance}
To evaluate the bound in Lemma \ref{le:TV-4}, we use again the change of measure trick, although it is more involved w.r.t. the one used for the relative entropty.

Define the tilted conditional distribution $P_{\overline{X}|\overline{Y}}$ and distribution $P_{\overline{Y}}$ via the following Radon-Nikodym derivatives,

\begin{align}
\frac{\mathrm{d}P_{\overline{X}|\overline{Y}}}{\mathrm{d}P_{X|Y}}(x,y)&:=\frac{\exp\left(\frac{\rho^*}{1-\rho^*}\imath(x;y)\right)}{\e\left[\exp\left(\frac{\rho^*}{1-\rho^*}\imath(X;Y)\right)|Y=y\right]}\label{eqn:mhym-b}\\
\frac{\mathrm{d}P_{\overline{Y}}}{\mathrm{d}P_{Y}}(y)	&:=\frac{\e^{{1-\rho^*}}\left[\exp(\frac{\rho^*}{1-\rho^*}\imath(X;Y))|Y=y\right]}{\e\left[\e^{1-\rho^*}\left[\exp(\frac{\rho^*}{1-\rho^*}\imath(X;Y))|Y\right]\right]},
\end{align}
where $\rho^*$ was defined in \eqref{eq:rho-defn}.
Also, for brevity let
\begin{equation}
	{\sf S}:={\e\left[\e^{1-\rho^*}\left[\exp(\frac{\rho^*}{1-\rho^*}\imath(X;Y))|Y\right]\right]}=\exp\left(\rho^* I_{\frac{1}{1-\rho^*}}(X;Y)\right),
\end{equation}
where $(X,Y)\sim P_{XY}$.

 We consider the cases $\rho^*<\frac{1}{2}$ and $\rho^*=1$, separately.

{\bf Case I:    $\rho^*<\frac{1}{2}$}. 
By Corollary \ref{cor:apx-tv} in Appendix \ref{apx:optimum} , $R$ and $\rho^*$ satisfy the following identity,
\begin{align}
R&=\frac{1}{1-\rho^*}\e[\imath_{X;Y}(\overline{X};\overline{Y})]~-\e\left[\log\e\left[\exp(\frac{\rho^*}{1-\rho^*}\imath(X;Y))|Y=\overline{Y}\right]	\right]\\
&=\e[ Z]\label{eqn:mhym-identity}
\end{align}
where the r.v. $Z$ is defined through,
\begin{align}
Z:=&\frac{1}{1-\rho^*}\imath_{X;Y}(\overline{X};\overline{Y})-\log\e\left[\exp(\frac{\rho^*}{1-\rho^*}\imath(X;Y))|Y=\overline{Y}\right].\label{eqn:TV-tilted}
\end{align}
Also for $k=1,\cdots,n$, let
\begin{align}
Z_k:=&\frac{1}{1-\rho^*}\imath_{X;Y}(\overline{X}_k;\overline{Y}_k)-\log\e\left[\exp(\frac{\rho^*}{1-\rho^*}\imath(X;Y))|Y=\overline{Y}_k\right],\label{eqn:mhym-e}
\end{align}
where $(\overline{X}_1,\overline{Y}_1),\cdots,(\overline{X}_n,\overline{Y}_n)$ are i.i.d and drawn from $P_{\overline{X},\overline{Y}}$. Now, we compute the expressions appeared in Lemma \ref{le:TV-4} in terms of $Z_1,\cdots,Z_n$. First, consider
\begin{align}
	\e[\exp(\imath(\bX;\bY))\mathbbm{1}\{(\bX,\bY)\in\mathcal{F}\}|\bY=\by]	&=\e\left[\exp (\frac{\rho^*}{1-\rho^*}\imath(\bX;\bY))\Big|\bY=\by\right]\nonumber\\&\qquad\qquad	\e\left[\exp(\frac{1-2\rho^*}{1-\rho^*}\imath_{\bX;\bY}(\overline{\bX};\overline{\bY}))\mathbbm{1}\{(\overline{\bX},\overline{\bY})\in\mathcal{F}\}\Big|\overline{\bY}=\by\right]\label{eqn:tv-c-m-1}\\
	&{=\e^{{2}{(1-\rho^*)}}\left[\exp (\frac{\rho^*}{1-\rho^*}\imath(\bX;\bY))\Big|\bY=\by\right]}\nonumber\\&~~~~~~~~~~~~~~~~{\e[\exp((1-2\rho^*)\sum_{k=1}^nZ_k)\mathbbm{1}\{(\overline{\bX},\overline{\bY})\in\mathcal{F}\}|\overline{\bY}=\by]},\label{eqn:tv-c-m-2}
\end{align}
where \eqref{eqn:c-o-m-1} follows by change of measure and \eqref{eqn:c-o-m-2} follows from the definition of $Z_k$. Similarly, we have
\begin{align}
	\e[\exp(3\imath(\bX;\bY))\mathbbm{1}\{(\bX,\bY)\in\mathcal{F}\}|\bY=\by]	&=\e[\exp(\frac{\rho^*}{1-\rho^*}\imath(\bX;\bY))|\bY=\by]\nonumber\\&	~~~~~~~~~~\e[\exp(\frac{3-4\rho^*}{1-\rho^*}\imath_{\bX;\bY}(\overline{\bX};\overline{\bY}))\mathbbm{1}\{(\overline{\bX},\overline{\bY})\in\mathcal{F}\}|\overline{\bY}=\by]\\
	&=\e^{{4}{(1-\rho^*)}}[\exp (\frac{\rho^*}{1-\rho^*}\imath(\bX;\bY))|\bY=\by]\nonumber\\&~~~~~~~~~~~\e[\exp((3-4\rho^*)\sum_{k=1}^nZ_k)\mathbbm{1}\{(\overline{\bX},\overline{\bY})\in\mathcal{F}\}|\overline{\bY}=\by].\label{eqn:tv-c-m-4}
\end{align}

Using \eqref{eqn:tv-c-m-2} and \eqref{eqn:tv-c-m-4}, the lower bound in Lemma \ref{le:TV-4} simplifies as follows,

\begin{align}
	&\e\left[\left(\dfrac{\e\left[\exp(3\imath(\mathbf{X};\mathbf{Y})\mathbbm{1}\{\mathbf{X},\mathbf{Y})\in\mathcal{F}\}|\bY\right]}{\e\left[\exp(\imath(\mathbf{X};\mathbf{Y})\mathbbm{1}\{\mathbf{X},\mathbf{Y})\in\mathcal{F}\}|\bY\right] ^{3}}+{3}\mu_n\e\left[\exp(\imath(\mathbf{X};\mathbf{Y})\mathbbm{1}\{\mathbf{X},\mathbf{Y})\in\mathcal{F}\}|\bY\right]^{-1} \right)^{-\frac{1}{2}}\right]
	\nonumber\\
	&=\e\Bigg[\e^{{1-\rho^*}}[\exp (\frac{\rho^*}{1-\rho^*}\imath(\bX;\bY))|\bY]
	\nonumber\\
	&\qquad\qquad\left(\dfrac{\e[\exp((3-4\rho^*)\sum_{k=1}^nZ_k)\mathbbm{1}\{(\overline{\bX},\overline{\bY})\in\mathcal{F}\}|\overline{\bY}=\bY]}{\e[\exp((1-2\rho^*)\sum_{k=1}^nZ_k)\mathbbm{1}\{(\overline{\bX},\overline{\bY})\in\mathcal{F}\}|\overline{\bY}=\bY] ^{3}}\right.
	\nonumber\\
	&\left.\qquad\qquad\qquad\qquad\qquad\qquad+\frac{3\mu_n}{\e[\exp((1-2\rho^*)\sum_{k=1}^nZ_k)\mathbbm{1}\{(\overline{\bX},\overline{\bY})\in\mathcal{F}\}|\overline{\bY}=\bY]}
\right)^{-\frac{1}{2}}\Biggm]
\label{eqn:1260-0}\\
&=\mathsf{S}^n\e\Bigg[\left(\dfrac{\e[\exp((3-4\rho^*)\sum_{k=1}^nZ_k)\mathbbm{1}\{(\overline{\bX},\overline{\bY})\in\mathcal{F}\}|\overline{\bY}]}{\e[\exp((1-2\rho^*)\sum_{k=1}^nZ_k)\mathbbm{1}\{(\overline{\bX},\overline{\bY})\in\mathcal{F}\}|\overline{\bY}] ^{3}}\right.
\nonumber\\
&\qquad\qquad\qquad\qquad\qquad\qquad\left.+\frac{6{\sf M}_n}{\e[\exp((1-2\rho^*)\sum_{k=1}^nZ_k)\mathbbm{1}\{(\overline{\bX},\overline{\bY})\in\mathcal{F}\}|\overline{\bY}]}
\right)^{-\frac{1}{2}}\Biggm]
\label{eqn:1260-1}\\
&={\sf S}^n\mathsf{M}_n^{-\rho^*}\e\Bigg[\left(\dfrac{\e[\exp((3-4\rho^*) \sum_{k=1}^n (Z_k-\e[Z]))\mathbbm{1}\{(\overline{\bX},\overline{\bY})\in\mathcal{F}\}|\overline{\bY}]}{\e[\exp((1-2\rho^*) \sum_{k=1}^n (Z_k-\e[Z]))\mathbbm{1}\{(\overline{\bX},\overline{\bY})\in\mathcal{F}\}|\overline{\bY}] ^{3}}\right.\nonumber\\&\left.\qquad\qquad\qquad\qquad+6\e[\exp((1-2\rho^*) \sum_{k=1}^n (Z_k-\e[Z]))\mathbbm{1}\{(\overline{\bX},\overline{\bY})\in\mathcal{F}\}|\overline{\bY}]^{-1}
\right)^{-\frac{1}{2}}\Biggr]\label{eqn:1260}
\end{align}
where 
\begin{itemize}
\item \eqref{eqn:tv-c-m-2} and \eqref{eqn:tv-c-m-4} yields \eqref{eqn:1260-0}
\item change of measure $P_Y\rightarrow P_{\overline{Y}}$ implies \eqref{eqn:1260-1}. Also, here $\mathsf{M}_n:=\exp(nR)$. 
\item The identity $R=\e[Z]$ gives \eqref{eqn:1260}.
\end{itemize}	

Now ${\sf S}^n{\sf M}_n^{-\rho^*}=\exp(-n\rho^*(R-I_{\frac{1}{1-\rho^*}}(X;Y)))$, which is the exponent appeared in Theorem \ref{thm:exact-tv}. So it is remained to bound the expectation inside \eqref{eqn:1260} to get the desired pre-factor. Let

\begin{align}
	\mathsf{P}=\e&\Bigg[\left(\dfrac{\e[\exp((3-4\rho^*) \sum_{k=1}^n (Z_k-\e[Z]))\mathbbm{1}\{(\overline{\bX},\overline{\bY})\in\mathcal{F}\}|\overline{\bY}]}{\e[\exp((1-2\rho^*) \sum_{k=1}^n (Z_k-\e[Z]))\mathbbm{1}\{(\overline{\bX},\overline{\bY})\in\mathcal{F}\}|\overline{\bY}] ^{3}}\right.\nonumber\\&\left.\qquad\qquad\qquad\qquad+6\e[\exp((1-2\rho^*) \sum_{k=1}^n (Z_k-\e[Z]))\mathbbm{1}\{(\overline{\bX},\overline{\bY})\in\mathcal{F}\}|\overline{\bY}]^{-1}
\right)^{-\frac{1}{2}}\Biggr],\label{eqn:EV-SF-P}
\end{align}

and set
\begin{align}
	\mathcal{F}:=\left\{-a\le\sum_{k=1}^n(Z_k-\e[Z])\le0\right\}.\label{eqn:mathcal-F-dfn}
\end{align}
where $a$ is a large enough fixed value, such that $\p[(\overline{X},\overline{Y})\in\mathcal{F}]\geq\frac{c}{\sqrt{n}}$ for some $c>0$. Then the pre-factor ${\sf P}$ can be lower-bounded as follows,
\begin{align}
	\mathsf{P}&\geq\e\left[\left(\frac{\mathsf{K}_1}{\p[\mathcal{F}|\overline{\bY}]^2}+\frac{\mathsf{K}_2}{\p[\mathcal{F}|\overline{\bY}]}\right)^{-\frac{1}{2}}\right]\\
	&\geq \e\left[\left(\frac{\mathsf{K}_1+K_2}{\p[\mathcal{F}|\overline{\bY}]^2}\right)^{-\frac{1}{2}}\right]\\
	&=\frac{1}{\sqrt{K_1+K_2}}\p[\mathcal{F}]\\
	&\geq \frac{C}{\sqrt{n}}\label{eqn:f-singular-f}
	\end{align}
where $K_1=\exp(3(1-2\rho^*)a)$, $K_2=6\exp((1-2\rho^*)a)$ and $C$ is a positive constant. Finally, \eqref{eqn:f-singular-f} completes the converse proof of Theorem \ref{thm:exact-tv} for the \underline{singular channels} with $\tau^*<1$.
\subsection*{\underline{Non-Singular channels:}}
The choice of $\mathcal{F}$ in \eqref{eqn:mathcal-F-dfn} led to  pre-factor scale $\frac{1}{\sqrt{n}}$  which is optimal for the singular channels. However there is a gap between it and the optimal pre-factor scale $n^{-\frac{1-\rho^*}{2}}$ for the non-singular channels. Here, we perturb the definition of $\mathcal{F}$ to get the optimal pre-factor scale.

Let
\begin{equation}
			\mathcal{F}:=
		\left\{ -(a+\frac{1}{2}\log n)\le\sum_{k=1}^{n}(Z_k-\e[Z])\le-\frac{1}{2}\log n\right\}
\end{equation}
and
\begin{equation}
			\mathcal{G}:=
		\left\{\by:
		\begin{aligned} 
		\left|\sum_{k=1}^n\e[Z_k|\overline{Y}_k=y_k]-n\e[Z]\right|&\le\sqrt{n}\\
		\left|\sum_{k=1}^n\mathrm{Var}[Z_k|\overline{Y}_k=y_k]-n\e[\mathrm{Var}[Z|\overline{Y}]]\right|&\le\frac{1}{2}n\e[\mathrm{Var}[Z|\overline{Y}]]\\
		\left|\sum_{k=1}^n\mathrm{M}_3[Z_k|\overline{Y}_k=y_k]-n\e[\mathrm{M}_3[Z|\overline{Y}]]\right|&\le\frac{1}{2}n\e[\mathrm{M}_3[Z|\overline{Y}]]
\end{aligned}
\right\}\label{eqn:violating}
\end{equation}
where for a r.v. $X$, $\mathrm{M}_3[X]\triangleq \e[|X-\e[X]|^3]$. The key property of non-singular channel is that $Z$ is not  a function of $Y$, a.s. $P_Y$. Since $P_Y\ll\gg P_{\overline{Y}}$, it is not also a function of $Y$, a.s. $P_{\overline{Y}}$. As a result, $\e[\mathrm{Var}[Z|\overline{Y}]]$ and $\e[\mathrm{M}_3[Z|\overline{Y}]]$ are strictly positive. 
\begin{lem}\label{le:lemma9}
For large enough $n$,

\begin{enumerate}
	\item There exists a constant $C_0$ such that for any $\by\in\mathcal{G}$, 
	\begin{equation}
\p\left[\mathcal{F}|\overline{\bY}=\by\right]\geq \frac{C_0}{\sqrt{n}}	
\end{equation}
\item There exists a constant $C_1$ such that
\begin{equation}\p[\overline{\bY}\in\mathcal{G}]\geq C_1.\end{equation}
\end{enumerate}	
\end{lem}
The proof of this lemma is relegated to the Appendix \ref{apx:lemma8}.

Using Lemma \ref{le:lemma9}, the pre-factor $\mathsf{P}$ in \eqref{eqn:EV-SF-P}  is lower-bounded as follows,
 \begin{align}
	\mathsf{P}&\geq\e\Bigg[\left(\dfrac{\e[\exp((3-4\rho^*) \sum_{k=1}^n (Z_k-\e[Z]))\mathbbm{1}\{(\overline{\bX},\overline{\bY})\in\mathcal{F}\}|\overline{\bY}]}{\e[\exp((1-2\rho^*) \sum_{k=1}^n (Z_k-\e[Z]))\mathbbm{1}\{(\overline{\bX},\overline{\bY})\in\mathcal{F}\}|\overline{\bY}] ^{3}}\right.\nonumber\\
	&\left.\qquad\qquad\qquad+6\e[\exp((1-2\rho^*) \sum_{k=1}^n (Z_k-\e[Z]))\mathbbm{1}\{(\overline{\bX},\overline{\bY})\in\mathcal{F}\}|\overline{\bY}]^{-1}
\right)^{-\frac{1}{2}}\mathbbm{1}\left\{\overline{\bY}\in\mathcal{G}\right\}\Biggr]\label{eqn:EV-SF-PP}\\
&\geq \e\left[\left(\frac{K_1n^{-\rho^*}}{\p[\mathcal{F}|\overline{\bY}]^2}+\frac{\mathsf{K}_2n^{\frac{1}{2}-\rho^*}}{\p[\mathcal{F}|\overline{\bY}]}\right)^{-\frac{1}{2}}\mathbbm{1}\left\{\overline{\bY}\in\mathcal{G}\right\}\right]\label{eqn:1340}\\
&\geq \frac{n^{-\frac{1-\rho^*}{2}}}{\sqrt{K_1C_0^{-2}+K_2C_0^{-1}}} \p\left[\overline{\bY}\in\mathcal{G}\right]\label{eqn:1350}\\
&\geq \frac{C_1}{\sqrt{K_1C_0^{-2}+K_2C_0^{-1}}}n^{-\frac{1-\rho^*}{2}}\label{eqn:1360}
\end{align}
where $K_1$ and $K_2$ were defined before, \eqref{eqn:1340} follows from the definition of $\mathcal{F}$, \eqref{eqn:1350} follows from the first item of Lemma \ref{le:lemma9} and \eqref{eqn:1360} follows from the second item of Lemma \ref{le:lemma9}.

Putting \eqref{eqn:1360} in \eqref{eqn:1260}, concludes the converse proof of Theorem \ref{thm:exact-tv} for the non-singular channels.

\subsection*{\bf Case II, $\rho^*=\frac{1}{2}$.}

It is shown in Appendix \ref{apx:optimum} that 
\begin{align}
	R&\ge\e[Z]\label{eqn:TV-tilt-c}
\end{align}
where $Z$ is defined by \eqref{eqn:TV-tilted} with $\rho^*=\frac{1}{2}$. 
Now set,
\begin{align}
&\mathcal{F}:=\left\{\sum_{k=1}^nZ_k
\le n\e[Z]\right\}
\end{align}
Then the lower-bound in Lemma \ref{le:TV-4} is lower-bounded as follows,
\begin{align}
	&\e\left[\left(\dfrac{\e\left[\exp(3\imath(\mathbf{X};\mathbf{Y})\mathbbm{1}\{\mathbf{X},\mathbf{Y})\in\mathcal{F}\}|\bY\right]}{\e\left[\exp(\imath(\mathbf{X};\mathbf{Y})\mathbbm{1}\{\mathbf{X},\mathbf{Y})\in\mathcal{F}\}|\bY\right] ^{3}}+{3}\mu_n\e\left[\exp(\imath(\mathbf{X};\mathbf{Y})\mathbbm{1}\{\mathbf{X},\mathbf{Y})\in\mathcal{F}\}|\bY\right]^{-1} \right)^{-\frac{1}{2}}\right]\nonumber\\
&=\mathsf{S}^n\e\Bigg[\left(\dfrac{\e[\exp(\sum_{k=1}^nZ_k)\mathbbm{1}\{(\overline{\bX},\overline{\bY})\in\mathcal{F}\}|\overline{\bY}]}{\e[\mathbbm{1}\{(\overline{\bX},\overline{\bY})\in\mathcal{F}\}|\overline{\bY}] ^{3}}+\frac{6{\sf M}_n}{\e[\mathbbm{1}\{(\overline{\bX},\overline{\bY})\in\mathcal{F}\}|\overline{\bY}]}
\right)^{-\frac{1}{2}}\Biggm]\label{eqn:similar-1}\\
&\ge{\sf S}^n\e\Bigg[\left(\dfrac{\exp(n\e[Z] )}{\p[(\overline{\bX},\overline{\bY})\in\mathcal{F}|\overline{\bY}] ^{2}}+\dfrac{6\exp(nR )}{\p[(\overline{\bX},\overline{\bY})\in\mathcal{F}|\overline{\bY}] }
\right)^{-\frac{1}{2}}\Biggr]\label{eqn:similar-2}\\
&\ge\frac{1}{\sqrt{7}}{\sf S}^n\exp(-n\frac{R}{2})\p[(\overline{\bX},\overline{\bY})\in\mathcal{F}]\label{eqn:similar-3}\\
&\ge C\exp\left(\frac{n}{2}(I_{{2}(X;Y)}-R)\right)\label{eqn:similar-4}
\end{align}
where 
\begin{itemize}
	\item \eqref{eqn:similar-1} follows from \eqref{eqn:1260-1} (which is valid for any $\rho^*$) with $\rho^*=\frac{1}{2}$,
	\item The definition of $\mathcal{F}$ gives \eqref{eqn:similar-2},
	\item $R\geq\e[Z]$ yields \eqref{eqn:similar-3},
	\item Berry-Essen approximation $\p[(\overline{\bX},\overline{\bY})\in\mathcal{F}]=\frac{1}{2}+O(\frac{1}{\sqrt{n}})$ results in \eqref{eqn:similar-4}.
\end{itemize}
Finally, this concludes the converse proof for $\rho^*=\frac{1}{2}$.
\section{Exact Analysis for the Achievability}\label{sec:achievability}
\subsection{Relative Entropy}
The starting point in the achievability proof is the following well-known upper bound on the relative entropy, ( which is an one-shot bound, see \cite[Appendix IV]{hayashi11} among many others),
\begin{equation}
	\e\left[ D\left( \mathsf{P}_{\mathbf{Y}}||P^{\otimes n}_{Y}\right) \right] \leq \e[\log 1+\mathsf{M}_n^{-1}\exp(\imath(\bX;\bY))]\label{eq:hayashi}
\end{equation}
where $\mathsf{M}_n:=\exp(nR)$. We know proceed to get an almost exact computable expression for the r.h.s. of \eqref{eq:hayashi}. To do this, we prove the following  general result,
\begin{thm}\it\label{thm:kl-up}
Let $(V_1,\cdots,V_n)$ be i.i.d. r.v.'s with $V_i\sim P_V$ and $\e[V]<0$. Further, assume that $V$ has finite moment generating function in the neighborhood of the origin. Let 
\begin{equation}
	\tau^*=\arg\max_{0\le\tau\le1} \log\e[\exp(\tau V)]
\end{equation}	
Then, if $\tau^*<1$, we have for some $C>0$ which does not depend on $n$ and depends only on $\tau^*$,
\begin{align}
\e\left[\log\left(1+\exp\left(\sum_{k=1}^nV_k\right)\right)\right]
\le\dfrac{C}{\sqrt{n}}	\e^n[\exp(\tau^*V)]
\end{align}
\end{thm}
\begin{rem}\it
	The previous technique \cite{hayashi11} for bounding the l.h.s. of \eqref{eq:hayashi} gives an upper bound with the same exponent but without the pre-factor $\frac{1}{\sqrt{n}}$.
\end{rem}

Setting $V_i\leftarrow (\imath_{X;Y}(X_i;Y_i)-R)$, in Theorem \ref{thm:kl-up} implies (notice that $\e[V]=I(X;Y)-R<0$),
\begin{align*}
	\e\left[ D\left( \mathsf{P}_{\mathbf{Y}}||P^{\otimes n}_{Y}\right) \right]=O\left(\frac{\exp(-n\tau^*R)}{\sqrt{n}}\e^n[\exp(\tau^*\imath(X;Y))]\right)
\end{align*}
where $\tau^*$ was defined in \eqref{eq:tau-defn}. This completes the proof of the achievability for the case $\tau^*<1$. The case $\tau^*=1$ is follows from the known result in \cite{hayashi11}. 

\begin{IEEEproof}[Proof of Theorem \ref{thm:kl-up}]
Define the tilted distribution $P_{\overline{V}}$ via the following Radon-Nikodym derivative,
\begin{equation}
\dfrac{\mathrm{d}P_{\overline{V}}}{\mathrm{d}P_{V}}(v):=\dfrac{\exp(\tau^*v)}{\e[\exp(\tau^*V)]} \triangleq \dfrac{\exp(\tau^*v)}{\mathsf{T}}
\end{equation}
if $\tau^*<1$, then the following equation  holds,
\begin{equation}
\e\left[~\overline{V}~\right]=\dfrac{\e[V\exp(\tau^*V)]}{\e[\exp(\tau^*V)]}	=\frac{d}{d\tau}\log\e[\exp(\tau V)]\Big|_{\tau^*}=0.\footnote{ It is worthy to mention that if $\e[V]>0$, then any tilted distribution with the positive tilted parameter $\tau$, has a positive mean. It follows from the convexity of the function $f(t)= \log\e[\exp(\tau V)]$ and the fact that $f'(t)=\e[V_{\tau}]$, where $V_{\tau}$ is a r.v. with the tilted distribution with the parameter $\tau$. }
\end{equation}
 Now we can write,
\begin{align}
&\e\left[\log\left(1+\exp\left(\sum_{k=1}^nV_k\right)\right)\right]\nonumber\\
&=\mathsf{T}^n\e\left[\exp(-\tau^*\sum_{k=1}^n\overline{V_k})\log\left(1+\exp\left(\sum_{k=1}^n\overline{V_k}\right)\right)\right]	\end{align}
where $(\overline{V}_1,\cdots,\overline{V}_n)$ are i.i.d. and distributed according to $P_{\overline{V}}$. The equality follows by change of measure.

\vskip 2mm

Let $S_n=\frac{1}{\sqrt{n}}\sum_{k=1}^n \overline{V_k}$ and $g(x):=\exp\left({-\tau^*}x\right)\log\left(1+\exp\left(x\right)\right)$. Then we have,
\begin{align}
&\e\left[\exp(-\tau^*\sum_{k=1}^n\overline{V_k})\log\left(1+\exp\left(\sum_{k=1}^n\overline{V_k}\right)\right)\right]\nonumber\\
	&=\e\left[\exp\left({-\tau^*\sqrt{n}}S_n\right)\log\left(1+\exp\left(\sqrt{n}S_n\right)\right)\right]\nonumber\\
	&=\int_{-\infty}^{\infty} g(\sqrt{n}x) dF_{S_n}(x)\\
&=g(\infty)F_{S_n}(\infty)-g(-\infty)F_{S_n}(-\infty)\nonumber\\
&\qquad\qquad-\sqrt{n}\int_{-\infty}^{\infty}F_{S_n}(x)g'(\sqrt{n} x)dx\label{eq:int-1}\\
&=-\sqrt{n}\int_{-\infty}^{\infty}F_{S_n}(x)g'(\sqrt{n} x)dx\label{eq:int-2}\\
&=-\dfrac{1}{\sqrt{n}}\int_{-\infty}^{\infty}\sqrt{n}F_{S_n}\left(\dfrac{x}{\sqrt{n}}\right)g'(x)dx\label{eq:int-3}
\end{align}
where \eqref{eq:int-1} follows by integration by part and \eqref{eq:int-2} is due to the fact that $g$ vanishes at  $\pm\infty$ (this is true, since $0<\tau^*<1$).

Let $\sigma^2:=\e\left[\overline{V}^2\right]$ and $\rho=\e\left[|\overline{V}|^3\right]$. By the Berry-Esseen theorem \cite[Theorem 9.8]{mitz},
\begin{equation}\sup _ { x } \left| F_{Y_{\sigma}} ( x ) - F _ {S_ n } ( x ) \right| \leq   \dfrac{3\rho}{  \sigma ^ { 3 } \sqrt { n } }\triangleq \dfrac{C_1}{  \sqrt { n } }
\end{equation}
where $F_{Y_{\sigma}} ( x )$ is the c.d.f. of a mean zero Gaussian random variable $Y _ { \sigma }$ with
variance $\sigma ^ { 2 } .$ Hence
\begin{align}
&\left|\int_{-\infty}^{\infty}\sqrt{n}F_{S_n}\left(\dfrac{x}{\sqrt{n}}\right)g'(x)dx\right|\\&\le \left|\int_{-\infty}^{\infty}\sqrt{n}F_{Y_\sigma}\left(\dfrac{x}{\sqrt{n}}\right)g'(x)dx\right|+{C_1}\int_{-\infty}^{\infty}|g'(x)|dx\\
&= \left|\int_{-\infty}^{\infty}\sqrt{n}\left(F_{Y_\sigma}\left(\dfrac{x}{\sqrt{n}}\right)-F_Y(0)\right)g'(x)dx\right|\nonumber\\
&\qquad\qquad\qquad\qquad\qquad\qquad\quad+{C_1}\int_{-\infty}^{\infty}|g'(x)|dx\label{eq:int-4}\\
&\le K\int_{-\infty}^{\infty}|xg'(x)|dx+{C_1}\int_{-\infty}^{\infty}|g'(x)|dx\label{eq:int-5}
\end{align}
where \eqref{eq:int-4} holds, since $\int_{-\infty}^\infty g'(x)dx=0$ (because $|g'|$ is integrable and again $g$ vanishes at infinities). Here, $K=\frac{1}{\sqrt{2\pi \sigma^2}}$ is the upper bound on $F'_{Y_\sigma}=f_{Y_\sigma}$. It is easy to verify that $g'(x)$ decays exponentially fast at $\pm\infty$. Thus both the integrals inside \eqref{eq:int-5} are convergent. In summary, we conclude that there exists a constant $C$ depending only on the distribution $P_V$, such that 
\begin{align}
\e\left[\exp\left({-\tau\sqrt{n}}S_n\right)\log\left(1+\exp\left(\sqrt{n}S_n\right)\right)\right]\le \dfrac{C}{\sqrt{n}}.
\end{align}	
\end{IEEEproof}

\subsection{TV-distance}
Theorem \ref{thm:TV-one-shot} implies achievability of the bound \eqref{eq:TV-Exact} without the pre-factor for any $\rho^*$. As a result, it yields the achievability  of the bound \eqref{eq:TV-Exact} for the case $\rho^*=\frac{1}{2}$. So it is only required to investigate the case $\rho^*<\frac{1}{2}$.

As in the proof of  Theorem \ref{thm:TV-one-shot}, we start with the following $n$-shot version of the  lower bound \eqref{eq:os-tv}, 
\begin{align}
	\e[\|\mathsf{P}_{\bY}-P_Y^{\otimes n}\|]\leq \mathbb{P}[\mathcal{F}^c]+\frac{1}{2}\e\left[\sqrt{\mathsf{M}_n^{-1}\e[\exp(\imath(\bX;\bY)){1 }\{(\bX,\bY)\in\mathcal{F}\}|\bY]}\right]\label{eq:nshot-tv}
\end{align}
where ${\sf M}_n=\exp(nR)$.

To compute this bound for the appropriate choice of $\mathcal{F}$ (will be determined later), we need the following lemma,
\begin{lem}[
{\cite[Lemma 47]{PPV2010}}
]\it\label{thm:BR-type}
Let $(V_1,\cdots,V_n)$ be i.i.d. zero mean r.v.'s with $V_i\sim P_V$. Further, assume that $V$ has finite third moment $T_3$ and non-zero variance $\sigma^2$. 
	Then, for any $A$,
\begin{align}
\e\left[\exp\left(- \sum_{k=1}^nV_k\right)\mathbbm{1}\left\{\sum_{k=1}^nV_k\ge A\right\}\right]
\le\dfrac{C}{\sqrt{n}}	\exp(-A)
\end{align}
where $C=\frac{2}{\sigma}(\frac{\log 2}{\sqrt{2\pi}}+\frac{12T_3}{\sigma^2})$.
\end{lem}

We investigate the non-singular channels and singular channels, separately.

\subsubsection{Non-singular channels}
Recall the definitions of $P_{\overline{X}|\overline{Y}}$, $P_{\overline{Y}}$, $\mathsf{S}$, $Z$, $Z_k$ and $\rho^*$ in the equations \eqref{eqn:mhym-b}—\eqref{eqn:mhym-e}. Also, let
\begin{equation}
	\mathcal{F}:=\left\{\sum_{k=1}^n Z_k\le -\frac{1}{2}\log n\right\}
\end{equation}
 We compute each term of \eqref{eq:nshot-tv} separately. 
First consider  


\begin{align}
\mathbb{P}[(\bX,\bY)\in\mathcal{F}^c]&=\mathsf{S}^n\e\bigg[\e^{{\rho^*}}[\exp(\frac{\rho^*}{1-\rho^*}\imath(\bX;\bY))|\bY=\overline{\bY}]\nonumber\\
	&~~~~~~~~~~	\e[\exp(-\frac{\rho^*}{1-\rho^*}\imath_{\bX;\bY}(\overline{\bX};\overline{\bY}))\mathbbm{1}\{(\overline{\bX},\overline{\bY})\notin\mathcal{F}\}|\overline{\bY}]\bigg]	\label{eqn:f-t-1}\\
	&={\sf S}^n\e\left[\exp(-{\rho^*}\sum_{k=1}^n Z_k)\mathbbm{1}\left\{\sum_{k=1}^n Z_k\ge n\e[Z]-\frac{1}{2}\log n\right\}\right]\label{eqn:f-t-2}\\
	&={\sf S}^n\mathsf{M}_n^{-\rho^*}{\e\left[\exp\left(-{\rho^*}\sum_{k=1}^n (Z_k- \e[Z])\right)\mathbbm{1}\left\{\sum_{k=1}^n (Z_k-\e[Z])\ge -\frac{1}{2}\log n\right\}\right]}\label{eqn:f-t-21}\\
	&\leq C_1{\sf S}^n\mathsf{M}_n^{-{\rho^*}}n^{-\frac{1-\rho^*}{2}}\label{eqn:f-t-3}
\end{align}
where 
\begin{itemize}
	\item change of measure $P_{XY}\rightarrow P_{\overline{X}\overline{Y}}$ implies \eqref{eqn:f-t-1},
	\item definitions of $Z_k$ and $\mathcal{F}$ gives \eqref{eqn:f-t-2},
   \item the identity $R=\e[Z]$ gives \eqref{eqn:f-t-21}
	\item  Lemma \ref{thm:BR-type} with $V_k\leftarrow \rho^*(Z_k-\e[Z])$ and $A\leftarrow-\frac{\rho^*}{2}\log n$ yields \eqref{eqn:f-t-3}.
\end{itemize} 

Next consider the second term of \eqref{eq:nshot-tv},
\begin{align}
&\e\left[\sqrt{\mathsf{M}_n^{-1}\e[\exp(\imath(\bX;\bY))\mathbbm{1}\{(\bX,\bY)\in\mathcal{F}\}|\bY]}\right]\label{eqn:s-t-00}\\
	&=\e\bigg[\mathsf{M}_n^{-\frac{1}{2}}\e^{{(1-\rho^*)}}\left[\exp (\frac{\rho^*}{1-\rho^*}\imath(\bX;\bY))\Big|\bY\right]\e^{\frac{1}{2}}[\exp((1-2\rho^*)\sum_{k=1}^nZ_k)\mathbbm{1}\{(\overline{\bX},\overline{\bY})\in\mathcal{F}\}|\overline{\bY}=\bY]
	\bigg]\label{eqn:s-t-0}\\
	&= {\sf S}^n\mathsf{M}_n^{-\frac{1}{2}}\e\left[\sqrt{\e\left[\exp((1-2\rho^*)\sum_{k=1}^n Z_k)\mathbbm{1}\left\{\sum_{k=1}^n Z_k\le n\e[Z]-\frac{1}{2}\log n\right\}\Bigg|\overline{\bY}\right]}~\right]\label{eqn:s-t-1}\\
	&\le {\sf S}^n\mathsf{M}_n^{-\frac{1}{2}}\sqrt{\e\left[\exp((1-2\rho^*)\sum_{k=1}^n Z_k)\mathbbm{1}\left\{\sum_{k=1}^n Z_k\le n\e[Z]-\frac{1}{2}\log n\right\}\right]}\label{eqn:s-t-2}\\
	&= {\sf S}^n\mathsf{M}_n^{-{\rho^*}}
	\sqrt{\e\left[\exp\left((1-2\rho^*)\sum_{k=1}^n (Z_k- \e[Z])\right)\mathbbm{1}\left\{\sum_{k=1}^n (Z_k-\e[Z])\le -\frac{1}{2}\log n\right\}\right]}\label{eqn:s-t-3}\\
	&\leq C_2{\sf S}^n\mathsf{M}_n^{-\rho^*}n^{-\frac{1-\rho^*}{2}}\label{eqn:s-t-4}
\end{align}
where 
\begin{itemize}
	\item putting \eqref{eqn:tv-c-m-2} in \eqref{eqn:s-t-00} yields \eqref{eqn:s-t-0},
	\item change of measure $P_{Y}\rightarrow P_{\overline{Y}}$ and the definitions of $Z_k$ and $\mathcal{F}$, imply \eqref{eqn:s-t-1},
	\item Jensen inequality for the concave mapping $x\mapsto \sqrt{x}$ gives \eqref{eqn:s-t-2},
	\item the identity $R=\e[Z]$ gives \eqref{eqn:s-t-3},
	\item  Lemma \ref{thm:BR-type} with $V_k\leftarrow -(1-2\rho^*)(Z_k-\e[Z])$ and $A\leftarrow\frac{1-2\rho^*}{2}\log n$ yields \eqref{eqn:s-t-4}.
\end{itemize}


Finally putting \eqref{eqn:f-t-3} and \eqref{eqn:s-t-4} together gives the desired result for the non-singular channels.
\subsubsection{Singular channel}
The Definition of singular channel implies that $\imath_{X;Y}(X;Y)$ is a function of $Y$, almost surely $P_Y$. For brevity, let $\imath_{X;Y}(x;y):=g(y)$. Then, the definitions of $P_{\overline{Y}}$ and $\mathsf{S}$ are reduced to
\begin{align}
\frac{\mathrm{d}P_{\overline{Y}}}{\mathrm{d}P_{Y}}(y)	&:=\frac{\exp\left({\rho^*}g(y)\right)}{\e\left[\exp\left({\rho^*}g(Y)\right)\right]}
\end{align}
\[
{\sf S}:=\e\left[\exp\left({\rho^*}g(Y)\right)\right]
\]
Also, the identity \eqref{eqn:mhym-identity} is reduced to,
\begin{align}
R&=\frac{1}{1-\rho^*}\e[\imath_{X;Y}(\overline{X};\overline{Y})]-\e\left[\log\e[\exp(\frac{\rho^*}{1-\rho^*}\imath(X;Y))|Y=\overline{Y}]	\right]\\
&=\e[g(\overline{Y})]
\end{align}
Set,\[\mathcal{F}:=\left\{\by:\imath(\bx;\by)\le n\e[g(\overline{Y})]\right\}=\left\{\by:\sum_{k=1}^ng(y_k)\le n\e[g(\overline{Y})]\right\},
\]
where $(\overline{Y}_1,\cdots,\overline{Y}_n)\sim P^{\otimes n}_{\overline{Y}}$. Now, we compute the terms inside \eqref{eq:nshot-tv}. First consider,
	\begin{align}
\mathbb{P}[(\bX,\bY)\in\mathcal{F}^c]&=\mathbb{P}\left[\sum_{k=1}^{n}g(Y_k)\ge n\e[g(\overline{Y})]\right]\\
&={\sf S}^n\e\left[\exp\left(-\rho^*\sum_{k=1}^{n}g(\overline{Y}_k)\right)\mathbbm{1}\left\{\sum_{k=1}^{n}g(\overline{Y}_k)\ge n\e[g(\overline{Y})]\right\}\right]\label{eqn:f-t-1-0}\\
	&={\sf S}^n\mathsf{M}_n^{-\rho^*}\e\left[\exp\left(-\rho^*(\sum_{k=1}^{n}g(\overline{Y}_k))-n\e[g(\overline{Y})]\right)\mathbbm{1}\left\{\sum_{k=1}^{n}g(\overline{Y}_k))\ge n\e[g(\overline{Y})]\right\}\right]\label{eqn:f-t-2-0}\\
	&\leq C_1\frac{{\sf S}^n\mathsf{M}_n^{-\rho^*}}{\sqrt{n}}\label{eqn:f-t-3-0}
\end{align}
where
	\begin{itemize}
	
	\item change of measure $P_{Y}\rightarrow P_{\overline{Y}}$ implies \eqref{eqn:f-t-1-0},
	
	\item the identity $R=\e[g(\overline{Y})]$ gives \eqref{eqn:f-t-2-0},
	\item  Lemma \ref{thm:BR-type} with $V_k\leftarrow -\rho^*(g(\overline{Y}_k)-\e[\overline{Y}])$ and $A=0$ yields \eqref{eqn:f-t-3-0}.
\end{itemize}

Next, consider
\begin{align}
&\e\left[\sqrt{\mathsf{M}_n^{-1}\e[\exp(\imath(\bX;\bY))\mathbbm{1}\{(\bX,\bY)\in\mathcal{F}\}|\bY]}\right]\nonumber\\
&=\e\left[\sqrt{\mathsf{M}_n^{-1}\e[\exp(\imath(\bX;\bY))\mathbbm{1}\{\imath(\bX;\bY)\le n\e[g(\overline{Y})]\}|\bY]}\right]\nonumber\\
&=\e\left[\sqrt{\mathsf{M}_n^{-1}\exp\left(\sum_{k=1}^{n}g(Y_k)\right)\mathbbm{1}\left\{\sum_{k=1}^{n}g(Y_k)\le n\e[g(\overline{Y})] \right\}}\right]\nonumber\\
	&={\sf S}^n\mathsf{M}_n^{-\frac{1}{2}}\e\left[\exp\left((\frac{1}{2}-\rho^*)\sum_{k=1}^{n}g(\overline{Y}_k)\right)\mathbbm{1}\left\{\sum_{k=1}^{n}g(\overline{Y}_k)\le n\e[g(\overline{Y})]\right\}\right]\label{eqn:s-t-1-0}\\
		&={\sf S}^n\mathsf{M}_n^{-\rho^*}\e\left[\exp\left((\frac{1}{2}-\rho^*)\left(\sum_{k=1}^{n}g(\overline{Y}_k)-\e[g(\overline{Y})])\right)\right)\mathbbm{1}\left\{\sum_{k=1}^{n}g(\overline{Y}_k)\le n\e[g(\overline{Y})] \right\}\right]\label{eqn:s-t-2-0}\\
	&\leq C_2\frac{{\sf S}^n\mathsf{M}_n^{-\rho^*}}{\sqrt{n}}\label{eqn:s-t-3-0}
	\end{align}
	where
	\begin{itemize}
	
	\item change of measure $P_{Y}\rightarrow P_{\overline{Y}}$ implies \eqref{eqn:s-t-1-0},
	
	\item the identity $R=\e[g(\overline{Y})]$ gives \eqref{eqn:s-t-2-0},
	\item  Lemma \ref{thm:BR-type} with $V_k\leftarrow -(\frac{1}{2}-\rho^*)(g(\overline{Y}_k)-\e[\overline{Y}])$ and $A=0$ yields \eqref{eqn:s-t-3-0}.
\end{itemize}

Finally putting \eqref{eqn:f-t-3-0} and \eqref{eqn:s-t-3-0} together gives the desired result for the singular channels.	
\appendices
\section{Proof of Lemma \ref{le:conv}}\label{apx:log}
Using the identity $u\log u+1-u=(u-1)^2\int_0^1 \frac{1-t}{1+t(u-1)}\mathsf{d}t$, we have 
\begin{IEEEeqnarray}{rCl}
\e\left[ U\log U\right] &=&\e\left[ U\log U+1-U\right] \\ &=&\e\left[ \left( U-1\right) ^{2}\int ^{1}_{0}\dfrac {1-t}{1+t\left( U-1\right) }dt\right]\\ 
& \ge &\dfrac{\left( \e\left[\displaystyle{\int} ^{1}_{0}\left( U-1\right) ^{2}\left( 1-t\right) dt\right] \right) ^{2}}{\e\left[ \displaystyle{\int} ^{1}_{0}\left( 1+t\left( U-1\right) \right) \left( U-1\right) ^{2}\left( 1-t\right) dt\right] }~~~~~\label{eq:cauchy}\\
& =&\dfrac {\dfrac {1}{4}\e\left[ \left( U-1\right) ^{2}\right] ^{2}}{\dfrac {1}{2}\e\left[ \left( U-1\right) ^{2}\right] +\dfrac {1}{6}\e\left[ \left( U-1\right) ^{3}\right] }\label{eq:kl-lower}
\end{IEEEeqnarray}
where \eqref{eq:cauchy} follows from Cauchy-Schwarz inequality.	
\section{Monotonicity of $\mathsf{L}_m$ and $\mathsf{V}_m$}\label{apx:mono-f-divergence}
We prove a more general result. Let $f:\mathbb{R}^{\ge0}\mapsto \mathbb{R}$ be a convex function with $f(1)=0$ and $D_f(P||Q)$ (defined below) is the $f$-divergence between $P$ and $Q$,
\[
D_f(P||Q):=\e\left[f\left(\dfrac{\mathsf{d}P}{\mathsf{d}Q}(Z)\right)\right]
\]
where $Z\sim Q$. 

Let 
$$\mathsf{L}_m^{(f)}:=\e\left[D_f\left(\mathsf{P}_Y^{(m)}||P_Y\right)\right],$$
 where $\mathsf{P}_Y^{(m)}(.):=\frac{1}{m}\sum_{k=1}^m P_{Y|X}(.|X(k))$, in which $(X_1,\cdots,X_m)\sim P_X\otimes\cdots\otimes P_X$.
\begin{lem}\label{le:mono-f-divergence}
$\mathsf{L}_m^{(f)}$ is a decreasing sequence of $m$.	
\end{lem}
\begin{IEEEproof}
Observe that 
\[
\dfrac{\mathsf{d}\mathsf{P}_Y^{(m)}}{\mathsf{d}P_Y}=\frac{1}{m}\sum_{k=1}^m\exp(\imath_{X;Y}(X(k);Y))
\]
Let 
$$Z_i:=\frac{1}{m-1}\sum_{k\neq i}\exp(\imath_{X;Y}(X(k);Y)).
$$
Then, we have 
\[
\dfrac{\mathsf{d}\mathsf{P}_Y^{(m)}}{\mathsf{d}P_Y}=\frac{1}{m}\sum_{i=1}^mZ_i\]
Using this and the Jensen inequality for the convex function $f$, we get 
\begin{align}
	\mathsf{L}_{m}^{(f)}&=\e\left[D_f\left(\mathsf{P}_Y^{(m)}||P_Y\right)\right]\\
	&=\e_{(X_1,\cdots,X_m,Y)\sim P_X\otimes\cdots\otimes P_X\otimes P_Y}\left[f\left(\dfrac{\mathsf{d}\mathsf{P}_Y^{(m)}}{\mathsf{d}P_Y}(Y)\right)\right]\\
             &=\e\left[f\left(\frac{1}{m}\sum_{i=1}^mZ_i\right)\right]\label{eq:j63}\\
             &\leq \frac{1}{m}\sum_{i=1}^m\e\left[f\left(Z_i\right)\right]\label{eq:j64}\\
             &=\e[f(Z_m)]\label{eq:j65}\\
             &=\e\left[f\left(\dfrac{\mathsf{d}\mathsf{P}_Y^{(m-1)}}{\mathsf{d}P_Y}(Y)\right)\right]\\
             &=\mathsf{L}_{m-1}^{(f)}
\end{align} 
where \eqref{eq:j64} follows from Jensen inequality and \eqref{eq:j65} follows from symmetry.	
\end{IEEEproof}
\section{Proof of Lemma \ref{le:2}}\label{apx:relative-entropy}
Consider,
\begin{align}
	\e\left[ \mathsf{L}_{M}\right] &=\e\left[  \left( \frac {1}{M}\sum ^{M}_{k=1}\exp\left( \imath\left( \bX_{k};\bY\right) \right) \right)\log\left( \frac {1}{M}\sum ^{M}_{k=1}\exp\left( \imath\left( \bX_{k};\bY\right) \right) \right)\right] \label{eq:PA-1}\\
	&\geq \e\left[  \left( \frac {1}{M}\sum ^{M}_{k=1}\exp\left( \imath\left( \bX_{k};\bY\right) \right) \right)\log\left( \frac {1}{M}\sum ^{M}_{k=1}\exp\left( \imath\left( \bX_{k};\bY\right) \right) \right) \mathbbm{1}\{M\le 2\mu_n\}\right]\label{eq:PA-2}\\
	&\geq \frac{1}{2\mu_n}\e\left[  \left( \sum ^{M}_{k=1}\exp\left( \imath\left( \bX_{k};\bY\right) \right) \right)\log\left( \frac {1}{M}\sum ^{M}_{k=1}\exp\left( \imath\left( \bX_{k};\bY\right) \right) \right) \mathbbm{1}\{M\le 2\mu_n\}\right]\label{eq:PA-3}\\
	&= \frac{1}{2\mu_n}\e\left[  \left( \sum ^{M}_{k=1}\exp\left( \imath\left( \bX_{k};\bY\right) \right) \right)\log\left( \frac {1}{M}\sum ^{M}_{k=1}\exp\left( \imath\left( \bX_{k};\bY\right) \right) \right) \right]\nonumber\\
	&\qquad\quad- \frac{1}{2\mu_n}\e\left[  \left( \sum ^{M}_{k=1}\exp\left( \imath\left( \bX_{k};\bY\right) \right) \right)\log\left( \frac {1}{M}\sum ^{M}_{k=1}\exp\left( \imath\left( \bX_{k};\bY\right) \right) \right) \mathbbm{1}\{M> 2\mu_n\}\right]\label{eq:PA-4}\\
	&=\frac{1}{2\mu_n}\e\left[  \left( \sum ^{M}_{k=1}\exp\left( \imath\left( \bX_{k};\bY\right) \right) \right)\log\left( \sum ^{M}_{k=1}\exp\left( \imath\left( \bX_{k};\bY\right) \right) \right) \right]-\frac{1}{2\mu_n}\e\left[  M\log M \right]\nonumber\\
	&\qquad\quad- \frac{1}{2\mu_n}\e\left[ M\mathsf{L}_M\mathbbm{1}\{M> 2\mu_n\}\right]\label{eq:PA-5}\\
		&=\frac{1}{2\mu_n}\e\left[  \left( \sum ^{M}_{k=1}\exp\left( \imath\left( \bX_{k};\bY\right) \right) \right)\log\left( \sum ^{M}_{k=1}\exp\left( \imath\left( \bX_{k};\bY\right) \right) \right) \right]-\frac{1}{2}\left( \log \mu_n+ \frac{1}{\mu_n} \right)\nonumber\\
	&\qquad\quad- \frac{1}{2\mu_n}\e\left[ M\mathsf{L}_M\mathbbm{1}\{M> 2\mu_n\}\right]\label{eq:PA-6}\\
	&=\frac{1}{2}\e\left[  \left( \frac{1}{\mu_n}
	\sum ^{M}_{k=1}\exp\left( \imath\left( \bX_{k};\bY\right) \right) \right)\log\left(\frac{1}{\mu_n} \sum ^{M}_{k=1}\exp\left( \imath\left( \bX_{k};\bY\right) \right) \right) \right]-\frac{1}{2\mu_n}\nonumber\\
	&\qquad\quad- \frac{1}{2\mu_n}\e\left[ M\mathsf{L}_M\mathbbm{1}\{M> 2\mu_n\}\right]\label{eq:PA-7}\\
	&=\frac{1}{2}\e\left[  \left( \frac{1}{\mu_n}
	\sum ^{M}_{k=1}\exp\left( \imath\left( \bX_{k};\bY\right) \right) \right)\log\left(\frac{1}{\mu_n} \sum ^{M}_{k=1}\exp\left( \imath\left( \bX_{k};\bY\right) \right) \right) \right]-\frac{1}{2\mu_n}\nonumber\\
	&\qquad\quad- \frac{1}{2\mu_n}\mathsf{L}_{\lceil2\mu_n\rceil}\e\left[ M\mathbbm{1}\{M\geq \lceil2\mu_n\rceil\}\right]\label{eq:PA-8}\\
	&=\frac{1}{2}\e\left[  \left( \frac{1}{\mu_n}
	\sum ^{M}_{k=1}\exp\left( \imath\left( \bX_{k};\bY\right) \right) \right)\log\left(\frac{1}{\mu_n} \sum ^{M}_{k=1}\exp\left( \imath\left( \bX_{k};\bY\right) \right) \right) \right]-\frac{1}{2\mu_n}\nonumber\\
	&\qquad\quad- \frac{1}{2}\mathsf{L}_{\lceil2\mu_n\rceil}\p\left[M\geq  2\mu_n-1\right]\label{eq:PA-9}
\end{align}	
where 
\begin{itemize}
	\item Identity \eqref{eq:PA-1} follows from the definition of relative entropy and the definition of $\mathsf{L}_M$ in \eqref{eq:LM-dfn},
	\item Equality \eqref{eq:PA-5} follows from the following identity,
	\begin{align}
		\e\left[  \left( \sum ^{M}_{k=1}\exp\left( \imath\left( \bX_{k};\bY\right) \right) \right)\log M \right]&=\e\left[\log M\e\left[   \sum ^{M}_{k=1}\exp\left( \imath\left( \bX_{k};\bY\right) \right) \Big|M \right]\right] \nonumber\\
		&=\e\left[ M  \log M\right] \label{eq:PAE}
	\end{align}, since 
	$\e[\exp( \imath( \bX_{k};\bY) )]=1$ for any $k$.
	\item Inequality \eqref{eq:PA-6} follows from the following inequality for the poisson r.v. $M$,
	\begin{align}
		\e[M\log M]=\e[M\log \mu_n]+\e\left[M\log \frac{M}{\mu_n}\right] \geq\mu_n\log\mu_n+\e\left[M\left(\frac{M}{\mu_n}-1\right)\right]=\mu_n\log\mu_n+1
	\end{align}
	where we used the inequality $\log x\leq x-1$, $\e[M^2]=\mu_n^2+\mu_n$ and $\e[M]=\mu_n$.
	\item Similar to \eqref{eq:PAE}, equality \eqref{eq:PA-7} follows from the following identity
	$$\e\left[   \frac{1}{\mu_n}
	\sum ^{M}_{k=1}\exp\left( \imath\left( \bX_{k};\bY\right) \right) \right]=\frac{1}{\mu_n}\e[M]=1$$
	\item \eqref{eq:PA-8} follows, because $\mathsf{L}_k$ is a decreasing sequence,
	\item Simple algebraic calculation for the Poisson r.v. $M$ implies \eqref{eq:PA-9}.
\end{itemize}
Finally, applying the following tail probability of the Poisson r.v. concludes the proof,
\begin{align}
	\p[M\ge 2\mu_n-1]\leq \p\left[M\ge \frac{3}{2}\mu_n\right] \le\varepsilon_{\frac{3}{2}}^{\mu_n}.\label{eq:PA-10}
\end{align}

\section{Proof of Lemma \ref{le:TV2} }\label{apx:TV}
The proof modifies the one given in \cite{yagli}.
\begin{IEEEproof}
\begin{align}
	\e\left[ \mathsf{V}_{M}\right] &=\frac{1}{2}\e\left[  \left| \frac {1}{M}\sum ^{M}_{k=1}\exp\left( \imath\left( \bX_{k};\bY\right) \right)-1 \right|\right] \label{eq:VA-1}\\
	&\geq \frac{1}{2}\e\left[  \left| \frac {1}{M}\sum ^{M}_{k=1}\exp\left( \imath\left( \bX_{k};\bY\right) \right)-1 \right| \mathbbm{1}\{M\le 2\mu_n\}\right]\label{eq:VA-2}\\
	&\geq \frac{1}{4\mu_n}\e\left[  \left| \sum ^{M}_{k=1}\exp\left( \imath\left( \bX_{k};\bY\right) \right)-M \right| \mathbbm{1}\{M\le 2\mu_n\}\right]\label{eq:VA-3}\\
	&= \frac{1}{4\mu_n}\e\left[  \left| \sum ^{M}_{k=1}\exp\left( \imath\left( \bX_{k};\bY\right) \right)-M \right| \right]- \frac{1}{4\mu_n}\e\left[  \left| \sum ^{M}_{k=1}\exp\left( \imath\left( \bX_{k};\bY\right) \right)-M \right| \mathbbm{1}\{M> 2\mu_n\}\right]\label{eq:VA-4}\\
	&=\frac{1}{4\mu_n}\e\left[  \left| \sum ^{M}_{k=1}\exp\left( \imath\left( \bX_{k};\bY\right) \right) -\mu_n\right| \right]-\frac{1}{4\mu_n}\e\left[  \left|M-\mu_n\right| \right]- \frac{1}{2\mu_n}\e\left[ M\mathsf{V}_M\mathbbm{1}\{M> 2\mu_n\}\right]\label{eq:VA-5}\\
		&\ge\frac{1}{4\mu_n}\e\left[  \left| \sum ^{M}_{k=1}\exp\left( \imath\left( \bX_{k};\bY\right) \right) -\mu_n\right| \right]-\frac{1}{4\sqrt{\mu_n}}- \frac{1}{2\mu_n}\e\left[ M\mathbbm{1}\{M> 2\mu_n\}\right]\label{eq:VA-6}\\
	&\ge\frac{1}{4}\e\left[  \left| \frac{1}{\mu_n}\sum ^{M}_{k=1}\exp\left( \imath\left( \bX_{k};\bY\right) \right) -1\right| \right]-\frac{1}{4\sqrt{\mu_n}}-\frac{1}{2} \varepsilon_{\frac{3}{2}}^{\mu_n}\label{eq:VA-7}
	\end{align}	
where 
\begin{itemize}
	\item Identity \eqref{eq:VA-1} follows from the definition of relative entropy and the definition of $\mathsf{V}_M$ in \eqref{eq:VM-dfn},
	\item Equality \eqref{eq:VA-5} follows from the triangle inequality and the  following identity,
	\begin{align}
		\e\left[  \left| \sum ^{M}_{k=1}\exp\left( \imath\left( \bX_{k};\bY\right) \right)-M \right| \mathbbm{1}\{M> 2\mu_n\}\right]
		&=\e\left[  \e\left[\left| \frac{1}{M}\sum ^{M}_{k=1}\exp\left( \imath\left( \bX_{k};\bY\right) \right)-1 \right|\Bigg| M \right] M\mathbbm{1}\{M> 2\mu_n\}\right]\nonumber\\
		&=2\e[M\mathsf{V}_M\mathbbm{1}\{M>2\mu_n\}] \label{eq:VAE}
	\end{align} 
	\item Inequality \eqref{eq:VA-6} follows from $\e[|X-\e[X]|]\le \mathrm{Var}[X]$ for any r.v. $X$,  $\mathrm{Var}[M]=\mu_n$ and $\mathsf{V}_k\le 1$ for any $k$.
	\item The inequality leading to \eqref{eq:VA-7} is currently proven in \eqref{eq:PA-9} and \eqref{eq:PA-10}.
	\end{itemize}

\end{IEEEproof}

\section{Proofs of the Lemma \ref{le:conv} and Lemma \ref{le:TV-conv}}
\subsection{Proof of Lemma \ref{le:conv}}\label{apx:sub-kl}
Write $T_1=\sum_{k=1}^MZ_k$, where $$Z_k:=\frac {1}{\mu _{n}}\exp \left( \imath\left( \mathbf{X}(k);\mathbf{Y}\right) \right) \mathbbm{1}\left\{ \left( \mathbf{X}\left( k\right) ,\mathbf{Y}\right) \in \mathcal{F}\right\}.$$
Utilizing Lemma \ref{le:conv} with $U\leftarrow T_1+1-\e[T_1|Y]$ in \eqref{eq:22}, yields	
\begin{align}
&\e[T\log T]\ge \e\Big[\e\Big[(T_1+1-\e[T_1|\mathbf{Y}])\log(T_1+1-\e[T_1|\mathbf{Y}])|\bY\Big]\Big]\\
&\ge   \e\left[\dfrac {\e\left[ \left(T_1-\e[T_1|\bY]\right) ^{2}|\mathbf{Y}\right] ^{2}}{{2}\e\left[\left(T_1-\e[T_1|\bY]\right) ^{2}|\mathbf{Y}\right] +\dfrac {2}{3}\e\left[ \left(T_1-\e[T_1|\bY]\right) ^{3}|\mathbf{Y}\right] }\right]\\
&= \e\left[\dfrac {\mu_n^2\e\left[Z_1^2|\mathbf{Y}\right] ^{2}}{{2\mu_n}\e\left[Z_1^2|\mathbf{Y}\right] +\dfrac {2}{3}\mu_n\e\left[ Z_1^{3}|\mathbf{Y}\right] }\right]\label{eq:simple-c}\\
&\ge \dfrac {\mu_n\e\left[Z_1^2\right] ^{2}}{{2}(\e\left[Z_1^2\right] +\dfrac {1}{3}\e\left[ Z_1^{3}\right]) }\label{eq:jen-x2y}\\
& \ge\frac{\mu_n}{4}\min\left\{ {\e\left[Z_1^2\right] },3\dfrac {\e\left[Z_1^2\right] ^{2}}{\e\left[ Z_1^{3}\right]) }\right\}\label{eq:29}
\end{align}
where 
\eqref{eq:simple-c} is a result of simple algebraic calculations using the moments of the Poisson r.v. $M$ and \eqref{eq:jen-x2y} follows by applying the Jensen inequality to the jointly convex function $f(x,y):=\frac{x^2}{y}$.

Next consider,
\begin{align}
	\e\left[Z_1^2\right]&=\frac{1}{\mu_n^2}\e\left[\exp(2\imath(\mathbf{X}(1);\mathbf{Y})\mathbbm{1}\{\mathbf{X}(1),\mathbf{Y})\in\mathcal{F}\}\right]\nonumber\\
	&=\frac{1}{\mu_n^2}\e\left[\exp(\imath(\mathbf{X};\mathbf{Y})\mathbbm{1}\{\mathbf{X},\mathbf{Y})\in\mathcal{F}\}\right];\label{eq:eq1}\\\e\left[Z_1^3\right]&=\frac{1}{\mu_n^3}\e\left[\exp(3\imath(\mathbf{X}(1);\mathbf{Y})\mathbbm{1}\{\mathbf{X}(1),\mathbf{Y})\in\mathcal{F}\}\right]\nonumber\\
	&=\frac{1}{\mu_n^3}\e\left[\exp(2\imath(\mathbf{X};\mathbf{Y})\mathbbm{1}\{\mathbf{X},\mathbf{Y})\in\mathcal{F}\}\right];\label{eq:eq2}
\end{align}
where the equalities \eqref{eq:eq1} and \eqref{eq:eq2} follow from change of measure, since $(\bX(1),\bY)\sim P^{\otimes n}_X P^{\otimes n}_Y$. Substituting \eqref{eq:eq1} and \eqref{eq:eq2} in \eqref{eq:29} concludes the proof.
\subsection{Proof of Lemma \ref{le:TV-conv}}\label{apx:sub-TV}
Utilizing Lemma \ref{le:TV-conv} with $U\leftarrow T_1$ in \eqref{eq:V22}, yields	
\begin{align}
\e[|T-1|]&\ge \e\left[\e\left[|T_1-\e[T_1|\mathbf{Y}]|\Big|\bY\right]\right]\\
&\ge   \e\left[\dfrac {\e\left[ \left(T_1-\e[T_1|\bY]\right) ^{2}|\mathbf{Y}\right] ^{3}}{\e\left[\left(T_1-\e[T_1|\bY]\right) ^{4}|\mathbf{Y}\right]  }\right]\\
&= \e\left[\dfrac {\mu_n^3\e\left[Z_1^2|\mathbf{Y}\right] ^{3}}{{\mu_n}\e\left[Z_1^4|\mathbf{Y}\right] +{3}\mu_n^2\e\left[ Z_1^{2}|\mathbf{Y}\right]^2 }\right]\label{eq:TV-simple-c}
\end{align}
where 
\eqref{eq:TV-simple-c} is a result of simple algebraic calculations using the moments of the Poisson r.v. $M$.
Next consider,
\begin{align}
	\e\left[Z_1^2|\bY\right]&=\frac{1}{\mu_n^2}\e\left[\exp(2\imath(\mathbf{X}(1);\mathbf{Y})\mathbbm{1}\{\mathbf{X}(1),\mathbf{Y})\in\mathcal{F}\}|\bY\right]\nonumber\\
	&=\frac{1}{\mu_n^2}\e\left[\exp(\imath(\mathbf{X};\mathbf{Y})\mathbbm{1}\{\mathbf{X},\mathbf{Y})\in\mathcal{F}\}|\bY\right];\label{eq:TV-eq1}\\
	\e\left[Z_1^4|\bY\right]&=\frac{1}{\mu_n^4}\e\left[\exp(4\imath(\mathbf{X}(1);\mathbf{Y})\mathbbm{1}\{\mathbf{X}(1),\mathbf{Y})\in\mathcal{F}\}|\bY\right]\nonumber\\
	&=\frac{1}{\mu_n^4}\e\left[\exp(3\imath(\mathbf{X};\mathbf{Y})\mathbbm{1}\{\mathbf{X},\mathbf{Y})\in\mathcal{F}\}\right|\bY];\label{eq:TV-eq2}
\end{align}
where the equalities \eqref{eq:TV-eq1} and \eqref{eq:TV-eq2} follows from change of measure, since $(\bX(1),\bY)\sim P^{\otimes n}_X P^{\otimes n}_Y$. Substituting \eqref{eq:TV-eq1} and \eqref{eq:TV-eq2} in \eqref{eq:TV-simple-c} concludes the proof.

{
\section{Proof of Lemma \ref{le:lemma9}}\label{apx:lemma8}

\subsection{ Proof of item 1:}
	Using the Berry-Essen theorem for the sum $\sum_{k=1}^n Z_k$, we get the following approximation for any $\by\in\mathcal{G}$,
\begin{align}
	\p\left[\mathcal{F}|\overline{\bY}=\by\right]&\geq \p\left[b_\by\le N\le c_\by\right]
	-6\dfrac{\sum_{k=1}^n\mathrm{M}_3[Z_k|\overline{Y}_k=y_k]}{\left(\sum_{k=1}^n\mathrm{Var}[Z_k|\overline{Y}_k=y_k]\right)^{\frac{3}{2}}}\\
	&\geq \p\left[b_\by\le N\le c_\by\right]
	-\frac{d}{\sqrt{n}}\label{eqn:BEEEE}
\end{align}
where $N$ is the standard normal random variable, 
$$b_\by=\frac{n\e[Z]-\sum_{k=1}^n\e[Z_k|\overline{Y}_k=y_k]-\frac{1}{2}\log n-a}{\sqrt{\sum_{k=1}^n\mathrm{Var}[Z_k|\overline{Y}_k=y_k]}}, ~~~c_\by=\frac{n\e[Z]-\sum_{k=1}^n\e[Z_k|\overline{Y}_k=y_k]-\frac{1}{2}\log n}{\sqrt{\sum_{k=1}^n\mathrm{Var}[Z_k|\overline{Y}_k=y_k]}},$$
and $d=\frac{9\sqrt{8}\e[\mathrm{M}_3[Z|\overline{Y}]]}{\e[\mathrm{Var}[Z|\overline{Y}]]^{\frac{3}{2}}}$.

Observe that for large enough $n$
$$\max\{|b_{\by}|,|c_\by|\}\leq \sqrt{\frac{2}{\e[\mathrm{Var}[Z|\overline{Y}]]}}\left(1+\frac{.5\log n+a}{\sqrt{n}}\right)\leq \frac{2}{\sqrt{\e[\mathrm{Var}[Z|\overline{Y}]]}}\triangleq \kappa.$$
Hence
\begin{equation}
\p\left[b_\by\le N\le c_\by\right]=\int_{b_\by}^{c_\by}\frac{\mathrm{e}^{-\frac{x^2}{2}}}{\sqrt{2\pi}}dx\geq (c_\by-b_\by)\frac{\mathrm{e}^{-\frac{\kappa^2}{2}}}{\sqrt{2\pi}}\geq  \frac{a\mathrm{e}^{-\frac{\kappa^2}{2}}}{\sqrt{3\pi\e[\mathrm{Var}[Z|\overline{Y}]]}}.\frac{1}{\sqrt{n}}\label{eqn:BE-N}
\end{equation}
Putting \eqref{eqn:BEEEE} and \eqref{eqn:BE-N} together yields that for large enough $a$,
\begin{equation}
\p\left[\mathcal{F}|\overline{\bY}=\by\right]\geq \frac{C}{\sqrt{n}}	
\end{equation}
for some constant $C$.

\subsection{ Proof of item 2:}
Using Chebyshev inequality, it is easy to show that the probability of violating the second constraint and the third constraint of \eqref{eqn:violating} is of order $O\left(n^{-.5}\right)$. Further, the Berry-Essen theorem implies that the probability of deviating of order $\sqrt{n}$ from the mean is lower-bounded by some non-zero constant $K$. Thus for large enough $n$, the probability of $\mathcal{G}$ is lower bounded by $\frac{K}{2}$. 

}

\section{Thinning property of a random Poisson sum}\label{apx:Poisson-tinning}
Let $X_1,X_2,\cdots$ be a sequence of i.i.d. random variables with the distribution $P_X$ and characteristic function $\varphi_X(t):=\e[\exp(\mathrm{i}tX)]$, where $X\sim P_X$. Let $M$ be a poisson r.v. with mean $\mu$ and  independent of the sequence $(X_1,X_2,\cdots)$. Further let $\mathcal{F}\subseteq\mathcal{X}$ be a measurable event w.r.t. to $P_X$. The following lemma is related to the \emph{thinning} property of Poisson random process \cite{last2017lectures}.
\begin{lem}
	Let $U=\sum_{k=1}^M X_k\mathbbm{1}\{X_k\in\mathcal{F}\}$ and $V=\sum_{k=1}^M X_k\mathbbm{1}\{X_k\notin\mathcal{F}\}$. Then $U$ and $V$ are independent. 
\end{lem}
\begin{IEEEproof}
	Let $\varphi_{U,V}(s,t):=\e[\exp(\mathrm{i}(sU+tV))]$, be the joint characteristic function of the pair $(U,V)$.  It suffices to show 
	\begin{equation}
	\varphi_{U,V}(s,t)=\varphi_{U}(s)\varphi_{V}(t),~~~~~~ \forall (s,t)\in\mathbb{R}^2.\label{thin-0}\end{equation}  
	The characteristic function of the random sum $U$ is given by (see \cite[Equation 2.4, P. 504]{Feller1971})
	\begin{align}
		\varphi_U(s)=\exp(\mu(\varphi_{X\mathbbm{1}\{X\in\mathcal{F}\}}(s)-1))
	\end{align} 
	Observe that 
	\begin{equation}
	\varphi_{X\mathbbm{1}\{X\in\mathcal{F}\}}(s)=\e[\exp(\mathrm{i}sX\mathbbm{1}\{X\in\mathcal{F}\})]=\p[X\notin\mathcal{F}]	+\e[\exp(\mathrm{i}sX)\mathbbm{1}\{X\in\mathcal{F}\}]
	\end{equation}
	Hence,
	\begin{align}
		\varphi_U(s)=\exp\left(\mu\big(\e[\exp(\mathrm{i}sX)\mathbbm{1}\{X\in\mathcal{F}\}]-\p[X\in\mathcal{F}]\big)\right)
	\end{align} 
	Similarly,
	\begin{align}
		\varphi_V(t)=\exp\left(\mu\big(\e[\exp(\mathrm{i}tX)\mathbbm{1}\{X\notin\mathcal{F}\}]-\p[X\notin\mathcal{F}]\big)\right)
	\end{align} 
	Thus,
	\begin{align}
	\varphi_{U}(s)\varphi_V(t)=\exp\left(\mu\big(\e[\exp(\mathrm{i}sX)\mathbbm{1}\{X\in\mathcal{F}\}]+\e[\exp(\mathrm{i}tX)\mathbbm{1}\{X\notin\mathcal{F}\}]-1\big)\right)\label{thin-1}
	\end{align} 
	Next consider,
	\begin{align}
		\varphi_{U,V}(s,t)&=\e[\exp(\mathrm{i}(sU+tV))]\\
		&=\e\left[\exp\left(\mathrm{i}\sum_{k=1}^M X_k(s\mathbbm{1}\{X_k\in\mathcal{F}\}+t\mathbbm{1}\{X_k\notin\mathcal{F}\})\right)\right]\\
		&=\exp\left(\mu(\varphi_{X(s\mathbbm{1}\{X\in\mathcal{F}\}+t\mathbbm{1}\{X\notin\mathcal{F}\})}(1)-1)\right)\label{thin-2}
	\end{align}
	where we have used again the formula \cite[Equation 2.4, P. 504]{Feller1971} for the Poisson random sum $\sum_{k=1}^M X_k(s\mathbbm{1}\{X_k\in\mathcal{F}\}+t\mathbbm{1}\{X_k\notin\mathcal{F}\})$.
	
	Now observe,
	\begin{align}
		\varphi_{X(s\mathbbm{1}\{X\in\mathcal{F}\}+t\mathbbm{1}\{X\notin\mathcal{F}\})}(1)&=\e\left[\exp(\mathrm{i}X(s\mathbbm{1}\{X\in\mathcal{F}\}+t\mathbbm{1}\{X\notin\mathcal{F}\}))\right]\\
		&=\e\left[\exp(\mathrm{i}sX)\mathbbm{1}\{X\in\mathcal{F}\}\right]+\e\left[\exp(\mathrm{i}tX)\mathbbm{1}\{X\notin\mathcal{F}\}\right]\label{thin-3}
	\end{align}
	Substituting \eqref{thin-3} in \eqref{thin-2} and comparing the result with \eqref{thin-1} yield \eqref{thin-0}. 
\end{IEEEproof}

\section{On the optimum values $\tau^*$ and $\rho^*$}\label{apx:optimum}
\begin{lem}The mapping $G:[0,1)\rightarrow \mathbb{R}$ defined by 
\begin{equation}
	G(\rho):=\log {\e\left[\e^{1-\rho}\left[\exp\left(\frac{\rho}{1-\rho}\imath_{X;Y}(X;{Y})\right)\Big|{Y}\right]\right]}
\end{equation} 
where $(X,{Y})\sim P_{XY}$, is convex.
\end{lem}
\begin{IEEEproof}
Let $F(\rho):={\e\left[\e^{1-\rho}\left[\exp\left(\frac{\rho}{1-\rho}\imath_{X;Y}(X;{Y})\right)\Big|{Y}\right]\right]}$. It suffices to show that for any $\theta,\alpha,\beta\in[0,1]$,
\begin{equation}
	F(\theta\alpha+\bar{\theta}\beta)\leq F(\alpha)^{\theta}F(\beta)^{\bar{\theta}},
\end{equation}
where $\bar{\theta}=1-\theta$.
The proof follows from repeatedly applying Jensen inequality, as follows,
\begin{align}
F(\alpha)^{\theta}F(\beta)^{\bar{\theta}}&=	\e^{\theta}\left[\e^{1-\alpha}\left[\exp\left(\frac{\alpha}{1-\alpha}\imath_{X;Y}(X;{Y})\right)\Big|{Y}\right]\right]\e^{\bar{\theta}}\left[\e^{1-\beta}\left[\exp\left(\frac{\beta}{1-\beta}\imath_{X;Y}(X;{Y})\right)\Big|{Y}\right]\right]\\
&\geq \e\left[\e^{\theta( 1-\alpha)}\left[\exp\left(\frac{\alpha}{1-\alpha}\imath_{X;Y}(X;{Y})\right)\Big|{Y}\right]\e^{\bar{\theta}(1-\beta)}\left[\exp\left(\frac{\beta}{1-\beta}\imath_{X;Y}(X;{Y})\right)\Big|{Y}\right]\right]\label{eqn:jen-con-1}\\
&= \e\left[\left(\e^{\frac{\theta( 1-\alpha)}{\theta( 1-\alpha)+\bar{\theta}(1-\beta)}}\left[\exp\left(\frac{\alpha}{1-\alpha}\imath_{X;Y}(X;{Y})\right)\Big|{Y}\right]\right.\right.\\&\qquad\qquad\qquad\qquad\left.\left.\e^{\frac{\bar{\theta}(1-\beta)}{\theta( 1-\alpha)+\bar{\theta}(1-\beta)}}\left[\exp\left(\frac{\beta}{1-\beta}\imath_{X;Y}(X;{Y})\right)\Big|{Y}\right]\right)^{\theta( 1-\alpha)+\bar{\theta}(1-\beta)}\right]\\
&\geq \e\left[\left(\e\left[\exp\left(\frac{\theta\alpha+\bar{\theta}\beta}{\theta( 1-\alpha)+\bar{\theta}(1-\beta)}\imath_{X;Y}(X;{Y})\right)\Big|{Y}\right]\right)^{\theta( 1-\alpha)+\bar{\theta}(1-\beta)}\right]\label{eqn:jen-con-2}\\
&=F(\theta\alpha+\bar{\theta}\beta)
\end{align}	
where \eqref{eqn:jen-con-1} and \eqref{eqn:jen-con-2} follow from the Holder inequality and the fact that  the mapping $x\mapsto x^{\theta( 1-\alpha)+\bar{\theta}(1-\beta)}$ is increasing.
\end{IEEEproof}
\begin{cor}
The function $H:[0,1)\mapsto \mathbb{R}$ defined by
\begin{align}
H(\rho)&=\frac{d}{d\rho}G(\rho):=\dfrac{1}{F(\rho)}
\e\left[\left\{\e^{1-\rho}\left[\exp\left(\frac{\rho}{1-\rho}\imath_{X;Y}(X;{Y})\right)\Big|{Y}\right]\right\}.\right.\nonumber\\
&\left.\left\{-\log\e\left[\exp\left(\frac{\rho}{1-\rho}\imath_{X;Y}(X;{Y})\right)\Big|{Y}\right]+\frac{1}{1-\rho}.\frac{\e\left[{\imath_{X;Y}(X;{Y})}\exp\left(\frac{\rho}{1-\rho}\imath_{X;Y}(X;{Y})\right)\Big|Y\right]}{\e\left[\exp\left(\frac{\rho}{1-\rho}\imath_{X;Y}(X;{Y})\right)\Big|Y\right]}
\right\}\right]	\label{eqn:Derivative}
\end{align}	
is increasing.
\end{cor}
Let $(X_{\rho},Y_\rho)$ be a pair of tilted random variables defined by the following pair of Radon-Nikodym derivatives,
\begin{align}
\frac{\mathrm{d}P_{{X}_\rho|{Y}_\rho}}{\mathrm{d}P_{X|Y}}(x,y)&:=\frac{\exp\left(\frac{\rho}{1-\rho}\imath(x;y)\right)}{\e\left[\exp\left(\frac{\rho}{1-\rho}\imath(X;Y)\right)|Y=y\right]}\\
\frac{\mathrm{d}P_{{Y}_\rho}}{\mathrm{d}P_{Y}}(y)	&:=\frac{\e^{{1-\rho}}\left[\exp(\frac{\rho}{1-\rho}\imath(X;Y))|Y=y\right]}{\e\left[\e^{1-\rho}\left[\exp(\frac{\rho}{1-\rho}\imath(X;Y))|Y\right]\right]}.
\end{align}
Also let $Z_\rho$ be a r.v. defined by
\begin{align}
Z_\rho:=&\frac{1}{1-\rho}\imath_{X;Y}({X}_\rho;{Y}_\rho)-\log\e\left[\exp(\frac{\rho}{1-\rho}\imath(X;Y))|Y={Y}_\rho\right]
\end{align}
Using these definitions, the r.h.s. of the derivative \eqref{eqn:Derivative}
can be  simplified as follows,
\begin{align}
	\dfrac{1}{F(\rho)}
\e&\left[\left\{\e^{1-\rho}\left[\exp\left(\frac{\rho}{1-\rho}\imath_{X;Y}(X;{Y})\right)\Big|{Y}\right]\right\}.\right.\nonumber\\
&\left.\left\{-\log\e\left[\exp\left(\frac{\rho}{1-\rho}\imath_{X;Y}(X;{Y})\right)\Big|{Y}\right]+\frac{1}{1-\rho}.\frac{\e\left[{\imath_{X;Y}(X;{Y})}\exp\left(\frac{\rho}{1-\rho}\imath_{X;Y}(X;{Y})\right)\Big|Y\right]}{\e\left[\exp\left(\frac{\rho}{1-\rho}\imath_{X;Y}(X;{Y})\right)\Big|Y\right]}
\right\}\right]	\nonumber\\
=\dfrac{1}{F(\rho)}
&\e\left[\left\{\e^{1-\rho}\left[\exp\left(\frac{\rho}{1-\rho}\imath_{X;Y}(X;{Y})\right)\Big|{Y}\right]\right\}.\right.\nonumber\\
&\left.\qquad\left\{-\log\e\left[\exp\left(\frac{\rho}{1-\rho}\imath_{X;Y}(X;{Y})\right)\Big|{Y}\right]+\frac{1}{1-\rho}.{\e\left[{\imath_{X;Y}(X_\rho;{Y}_\rho)}\Big|Y_\rho=Y\right]}\right\}\right]\label{eqn:c-o-m-1}	\\
=& \e\left[-\log\e\left[\exp\left(\frac{\rho}{1-\rho}\imath_{X;Y}(X;{Y})\right)\Big|{Y}=Y_\rho\right]+\dfrac{1}{1-\rho}.{\e\left[{\imath_{X;Y}(X_\rho;{Y}_\rho)}\Big|Y_\rho\right]}\right]\label{eqn:c-o-m-2}\\
=&\e[Z_\rho]	
\end{align}
where \eqref{eqn:c-o-m-1} and \eqref{eqn:c-o-m-2} follow from the change of measures $P_{X|Y}\rightarrow P_{X_\rho|Y_\rho}$ and $P_Y\rightarrow P_{Y_\rho}$, respectively.

Now let $R>I(X;Y)$ be a fixed number and define,
\begin{equation}
	\rho^*:=\arg\max_{0\leq\rho\leq\frac{1}{2}} \rho\left(R-I_{\frac{1}{1-\rho}(X;Y)}\right)=\arg\max_{0\leq\rho\leq\frac{1}{2}} \rho R-G(\rho)
\end{equation}
Since $G$ is convex, the function $f(\rho):=\rho R-G(\rho)$ gets its maximum either at the end-points $0,\frac{1}{2}$ or an interior point $\rho^*$ of the interval $[0,\frac{1}{2}]$ such that $\rho^*=H(\rho^*)=\e[Z_{\rho^*}]$. Also $\rho^*\neq 0$, because $f'(0)=R-I(X;Y)>0$. Therefore, if $\rho^*$ is not an interior point, it should be $\rho^*=\frac{1}{2}$ and it happens, if $R\geq\e[Z_{\frac{1}{2}}]$ due to the fact that $\e[Z_\rho]=H(\rho)$ is increasing. In summary, we proved the following,
\begin{cor}\label{cor:apx-tv}
We have,
\begin{equation}
	\rho^*=\left\{\begin{array}{lr}
		\frac{1}{2}& R\geq\e[Z_{\frac{1}{2}}]\\
		t<\frac{1}{2}& R=\e[Z_t]
	\end{array}\right.
\end{equation}	
\end{cor}
 
 A similar argument shows the following counterpart for the optimization in \eqref{eq:tau-defn}.
 \begin{lem}\label{le:apx-kl}
 We have 
 \begin{equation}
	\tau^*=\left\{\begin{array}{lr}
		1& R\geq\e[\imath_{X;Y}(X_1;Y_1)]\\
		t<1& R=\e[\imath_{X;Y}(X_t;Y_t)]
	\end{array}\right.
\end{equation}	
where $(X_\tau,Y_\tau)$ is defined via the following Radon-Nikodym derivative,
\begin{equation}
\dfrac{\mathrm{d}P_{X_\tau Y_\tau}}{\mathrm{d}P_{XY}}(x,y):=\dfrac{\exp(\tau\imath_{X;Y}(x;y))}{\e[\exp(\tau\imath_{X;Y}(X;Y))]} 
\end{equation}
 \end{lem}

%
%
%
%
%
%
%

\bibliographystyle{unsrt}
\bibliography{Ref}

\end{document}